\documentclass[a4paper,12pt]{article}
\pdfoutput=1
\usepackage{epsfig}
\usepackage{amssymb}
\usepackage{amsfonts}
\usepackage{amsmath}
\usepackage{euscript}
\usepackage{verbatim}
\usepackage{latexsym}
\usepackage{graphicx}
\usepackage{caption}
\usepackage{float}
\usepackage{subcaption}

\usepackage{listings}

\usepackage{booktabs}

\newif\ifdtup

\catcode`\@=11

\@addtoreset{equation}{section}

\def\@normalsize{\@setsize\normalsize{15pt}\xiipt\@xiipt
\abovedisplayskip 14pt plus3pt minus3pt%
\belowdisplayskip \abovedisplayskip
\abovedisplayshortskip \z@ plus3pt%
\belowdisplayshortskip 7pt plus3.5pt minus0pt}

\def\small{\@setsize\small{13.6pt}\xipt\@xipt
\abovedisplayskip 13pt plus3pt minus3pt%
\belowdisplayskip \abovedisplayskip
\abovedisplayshortskip \z@ plus3pt%
\belowdisplayshortskip 7pt plus3.5pt minus0pt
\def\@listi{\parsep 4.5pt plus 2pt minus 1pt
     \itemsep \parsep
     \topsep 9pt plus 3pt minus 3pt}}

\relax

\catcode`@=12

\catcode`\@=11

\def\section{\@startsection{section}{1}{\z@}{3.5ex plus 1ex minus
   .2ex}{2.3ex plus .2ex}{\large\bf}}

\def\SymBoxes#1#2#3#4{\newdimen\un@t \un@t#3%
\raisebox{#1}{\rule{#2\un@t}{#4}\hskip-#2\un@t% lower horizontal
\@tempdimb\un@t \advance\@tempdimb by-#4\@tempcntb#2\relax%
\@whilenum{\@tempcntb>0}\do{%                         % #2 vertical lines
\rule{#4}{\un@t}\hskip\@tempdimb \advance\@tempcntb by\m@ne}%
\hskip-#2\un@t \rule[\un@t]{#2\un@t}{#4}%
\rule[\un@t]{#4}{#4}\hskip-#4%             % upper horizontal line
\rule{#4}{\un@t}}\hskip-#4}                % rightest vertical line
\begin{document}
%\begin{letter}{~}

%%%%%%Define some new commands and  macros
\newcommand{\beq}{\begin{equation}}
\newcommand{\eeq}{\end{equation}}
\newcommand{\bea}{\begin{eqnarray}}
\newcommand{\eea}{\end{eqnarray}}
\newcommand{\beas}{\begin{eqnarray*}}
\newcommand{\eeas}{\end{eqnarray*}}
\newcommand{\defi}{\stackrel{\rm def}{=}}
\newcommand{\non}{\nonumber}
\newcommand{\bquo}{\begin{quote}}
\newcommand{\enqu}{\end{quote}}
%%%%%%%%%%%%%%%%
\renewcommand{\(}{\begin{equation}}
\renewcommand{\)}{\end{equation}}
%%%%%%%%%%%%%%%%%%%%%%%%%%%%%%%%%% definitions
\def \eqn#1#2{\begin{equation}#2\label{#1}\end{equation}}

\def\e{\epsilon}
\def\IZ{{\mathbb Z}}
\def\IR{{\mathbb R}}
\def\IC{{\mathbb C}}
\def\IQ{{\mathbb Q}}
\def\de{\partial}
\def\Tr{ \hbox{\rm Tr}}
\def\H{ \hbox{\rm H}}
\def\HE{ \hbox{$\rm H^{even}$}}
\def\HO{ \hbox{$\rm H^{odd}$}}
\def\K{ \hbox{\rm K}}
\def\Im{ \hbox{\rm Im}}
\def\Ker{ \hbox{\rm Ker}}
\def\const{\hbox {\rm const.}}
\def\o{\over}
\def\im{\hbox{\rm Im}}
\def\re{\hbox{\rm Re}}
\def\bra{\langle}\def\ket{\rangle}
\def\Arg{\hbox {\rm Arg}}
\def\Re{\hbox {\rm Re}}
\def\Im{\hbox {\rm Im}}
\def\exo{\hbox {\rm exp}}
\def\diag{\hbox{\rm diag}}
\def\longvert{{\rule[-2mm]{0.1mm}{7mm}}\,}
\def\a{\alpha}
\def\dag{{}^{\dagger}}
\def\tq{{\widetilde q}}
\def\p{{}^{\prime}}
\def\W{W}
\def\N{{\cal N}}
\def\hsp{,\hspace{.7cm}}

\def\br{\nonumber}
\def\IZ{{\mathbb Z}}
\def\IR{{\mathbb R}}
\def\IC{{\mathbb C}}
\def\IQ{{\mathbb Q}}
\def\IP{{\mathbb P}}
\def \eqn#1#2{\begin{equation}#2\label{#1}\end{equation}}

\newcommand{\C}{\ensuremath{\mathbb C}}
\newcommand{\Z}{\ensuremath{\mathbb Z}}
\newcommand{\R}{\ensuremath{\mathbb R}}
\newcommand{\rp}{\ensuremath{\mathbb {RP}}}
\newcommand{\cp}{\ensuremath{\mathbb {CP}}}
\newcommand{\vac}{\ensuremath{|0\rangle}}
\newcommand{\vact}{\ensuremath{|00\rangle}                    }
\newcommand{\oc}{\ensuremath{\overline{c}}}
\newcommand{\psizero}{\psi_{0}}
\newcommand{\phizero}{\phi_{0}}
\newcommand{\hzero}{h_{0}}
\newcommand{\psiin}{\psi_{\rh}}
\newcommand{\phiin}{\phi_{\rh}}
\newcommand{\hin}{h_{\rh}}
\newcommand{\rh}{r_{h}}
\newcommand{\rb}{r_{b}}
\newcommand{\psibnd}{\psi_{0}^{b}}
\newcommand{\psibndp}{\psi_{1}^{b}}
\newcommand{\phibnd}{\phi_{0}^{b}}
\newcommand{\phibndp}{\phi_{1}^{b}}
\newcommand{\gbnd}{g_{0}^{b}}
\newcommand{\hbnd}{h_{0}^{b}}
\newcommand{\zh}{z_{h}}
\newcommand{\zb}{z_{b}}
\newcommand{\man}{\mathcal{M}}
\newcommand{\hbr}{\bar{h}}
\newcommand{\tbr}{\bar{t}}

\begin{titlepage}
%\begin{flushright} CHEP XXXXX
%ULB-TH/09-10\\
%hep-th/yymmnnn\\ \end{flushright}
%\bigskip

\def\thefootnote{\fnsymbol{footnote}}

\begin{center}
{\large
{\bf A Central Charge for sub-AdS Holography} \\ 
\vspace{0.2cm}
%{\bf Submatrices, Long Strings and Small Black Holes}
}

\end{center}

\vspace{0.2cm}

%\bigskip
\begin{center}
Abir Ghosh$^a$\footnote{\texttt{abirghosh.physics@gmail.com}}, \ Chethan Krishnan$^a$\footnote{\texttt{chethan.krishnan.physics@gmail.com}}, \  Adinath Mahalingam$^a$\footnote{\texttt{adiphy2002@gmail.com}}
%\vspace{0.1in}
\end{center}

\renewcommand{\thefootnote}{\arabic{footnote}}

\begin{center}
%\vspace{0.2cm}

$^a$ {Center for High Energy Physics,\\
Indian Institute of Science, Bangalore 560012, India}\\

\end{center}
\vspace{-0.15in}
\noindent
\begin{center} {\bf Abstract} \end{center}
We synthesize and sharpen various observations about sub-AdS holography in the literature to associate a central charge to sub-AdS scales of AdS$ \times X$ geometries. A key ingredient in our proposal is the idea that the AdS length is the length of the long string in a stack of $N$ {\em backreacted} D-branes. This allows us to make statements about sub-AdS scales by considering long strings in sub-stacks of $M < N$ branes. Our proposal applies in general dimensions, connects with and refines previous ideas about sub-matrix deconfinement \& long strings, makes crucial use of the compact space, and is consistent with the expected thermodynamics of small black holes localized on $X$. Some of our arguments draw intuition from separating branes, and have natural connections to (heating up) the Coulomb branch of the gauge theory. We apply related ideas to non-conformal D-branes at zero and finite temperature, discuss the holographic bound and IR/UV duality in theories with 16 supercharges, and find broad consistency.

\vspace{1.6 cm}
\vfill

\end{titlepage}

\setcounter{footnote}{0}

\tableofcontents

%%%%%%%%%%%%%%%%%%%%%%%%%%%%%%%%%%%%%%%%%%%%%%%%%%%%%%%%%%%%%%%%%%%%%%%%%%%%%%%%%%%%%%%%%%%%%%
%%%%%%%%%%%%%%%%%%%%%%%%%%%%%%%%%%%%%%%%%%%%%%%%%%%%%%%%%%%%%%%%%%%%%%%%%%%%%%%%%%%%%%%%%%%%%%
\section{Introduction}

%Quantum gravity in asymptotically AdS space is believed to be a conformal gauge theory. 
Much of the work on the AdS/CFT correspondence \cite{Maldacena} focuses on scales larger than the AdS length scale \cite{WittenAdSCFT}. In the bulk, this allows one to ignore the details of the compactification. On the boundary, a dual of this fact is that one can learn quite a bit, just from conformal invariance\footnote{Note that this includes CFTs on backgrounds which break conformal invariance, where we are interested in states whose relevant scales are bigger than the scale of the background. Large black holes in global AdS can be studied by placing the CFT on a cylinder, at temperatures larger than the radius of the cylinder. In global AdS, the holographic RG flow is that of a conformal field theory in the UV of the field theory, but not below the AdS scale.}. Conformal invariance is realized as the isometry of the AdS part of the geometry, which comes into full effect at super-AdS scales. This is the IR of the bulk.  At shorter length scales, the bulk geometry of AdS$_5 \times S^5$ is effectively a 10-dimensional flat space, and boundary conformal invariance is less powerful in making useful statements. In understanding sub-AdS scales therefore, one expects that (a) the compact space of the bulk geometry is likely to play an important role, and (b) more detailed aspects of the gauge theory beyond its conformal invariance will be significant.

Despite its relative lack of footprint in current research\footnote{But see \cite{Karch, Trivedi} and related papers for a line of recent work that is closely related. }, there are a few reasons why sub-AdS holography will be the focus of this paper. The first is that in order to understand holography in flat space (or in an accelerating universe), we will have to let go of the safety of the AdS length scale. Although much work has been done on asymptotic symmetries and related ideas in flat space, many of the key conceptual questions have remained unanswered. There is a sense in which flat space is naturally treated as an open system -- black holes evaporate away completely and signals can be ``lost" into the asymptotic region. AdS on the other hand is naturally viewed as a box to contain gravity, in which black holes can be stable and the natural boundary condition sends signals back into the bulk in finite time. 

A paradigmatic illustration of the distinction between sub- and super-AdS scales is provided by small vs. large black holes in global AdS$_5 \times S^5$. Large-AdS black holes have a simple and beautiful understanding in terms of the high-temperature deconfined phase of the large-$N$ CFT on a cylinder \cite{WittenBH}. But the small black hole\footnote{Sometimes the phrase small black hole is used to describe black holes which are small compared to the AdS scale in an AdS geometry without reference to a compact space. This fails to distinguish between a black brane smeared on the $S^5$ vs a black hole that is localized on the $S^5$. The former suffers from a Gregory-Laflamme instability and is not a long-lived object from the dual CFT perspective. In this paper, by small black hole we will always mean a black hole that is localized on the $S^5$, but we will have more to say about both these objects later in the paper.}, which is a black hole small enough to be localized on the $S^5$, is qualitatively different. It is never the dominant thermodynamic phase and it has negative specific heat, and therefore it radiates away. A small black hole with a horizon radius hierarchically lower than the AdS length scale is essentially a 10-dimensional flat space black hole. While the large AdS black hole can be viewed as a coarse-grained description of a heavy eigenstate of the CFT, the small AdS black hole is an unstable (but long-lived) state. Precisely because of this, this is the setting in AdS/CFT where we can hope to learn about flat-space black holes and their Page curves, and more broadly about the general mechanisms underlying flat-space holography.

Our goal in this paper is to take some small steps towards concretizing sub-AdS holography, slightly refining older ideas in \cite{Verlinde, Hanada}. A key perspective we will adopt is that we will think of the AdS length scale as the length of the long string in a stack of $N$ D-branes that have backreacted\footnote{We will introduce these ideas in more detail, as we proceed.}.  Together with the beautiful and simple fact that a constant overall rescaling of an AdS $\times S$ metric simply results in a rescaling of the AdS length associated to the metric, we will immediately be led to claims about sub-AdS scales via long strings in sub-stacks of $M$ D-branes. Our observations slightly modify and generalize some of the suggestions in \cite{Verlinde}. We will find further support for our arguments by considering the Coulomb branch of the gauge theory at finite temperature, and connecting it to localized black holes. We will conclude by discussing Dp-branes at zero and finite temperature and discussing the holographic bound (and related physics) in them.

The results of this paper strengthen the heuristic picture that the nature of the radial coordinate in an AdS $\times S$ geometry changes, as one passes through the AdS length scale. On super-AdS scales, all the matrix degrees of freedom of the $N$ D-branes are accessible, and the radial coordinate is simply a way to keep track of the UV cut-off of the theory. This is indicated by the fact that the central charge of the theory does not change once we are above the AdS length scale. But below the AdS length, the bulk radial coordinate knows about the number of matrix degrees of freedom that are accessible -- the larger the relevant scale, the bigger the sub-stack of branes that are responsible for the dynamics. Correspondingly, the central charge scales in a precise manner.

\section{Substacks, Submatrices and Backreacted Long Strings} 

 The intuitive picture that will be guiding a lot of our discussion is that {\em the AdS length scale is the length scale associated to the long string in a stack of $N$ backreacted D-branes}\footnote{Closely related ideas have appeared before -- see in particular van Leuven-Visser-Verlinde \cite{Verlinde}, which is closest to our work. As far as we are aware, the earliest source of the idea that small black holes in AdS should be understood in terms of submatrices appears in the work of Asplund and Berenstein \cite{Berenstein}. The importance of submatrices in QCD and AdS/CFT has been discussed in a series of thought-provoking papers by Hanada and collaborators, we list some that we have found useful: \cite{Hanada1, Hanada2, Hanada3, Hanada4}.}.
 Note that when $N$ is large and the branes are heavy enough to backreact, this is the system that produces the AdS throat. 
    
Let us start with some basic observations. The first is that in the derivation of the usual AdS/CFT correspondence \cite{Maldacena}, the decoupling limit involves taking $r$ and $\alpha'$ to zero, while holding $r/\alpha'$ fixed, where $r$ is the radial coordinate of the original supergravity brane solution. Note that $U \equiv r/\alpha'$ here is dimensionful. In practice this means that the metric measured in units of $\alpha' (=l_s^2)$ stays fixed in the near-horizon limit.  We will write this as: 
\bea
ds^2 = l_s^2\left(\frac{U^2}{\sqrt{4 \pi g_s N}}(-dt^2+d{\bf x}^2)+\sqrt{4 \pi g_s N}\frac{dU^2}{U^2}+\sqrt{4 \pi g_s N}d\Omega_5^2\right) \label{NHmetric}
\eea
We are phrasing the present discussion in the familiar context of AdS$_5 \times S^5$, but the ideas are quite general and we will discuss some other examples later.  As well known, the resulting AdS length scale is
\bea
L^4 = 4 \pi g_s l_s^4 N \label{LvsN} 
\eea
We will view this as the backreacted long string length scale in a stack of $N$ D-branes. %We will call the $N=1$ version of this metric, the string scale AdS$_5 \times S^5$ metric. We will view it as an auxiliary mathematical object, and it is {\em not} to be interpreted as a classical geometric quantity. 
Classical geometry is meaningful around $N=\infty$. The results of \cite{Banks, Rajaraman, Mazzucato} show that the AdS$_5 \times S^5$ length does not get corrections in $\alpha'$ to any order, in the large-$N$ limit. When we are studying substacks of $M$ D-branes, we will implicitly assume that we are working with $M, N \rightarrow \infty$ with $M/N$ held fixed.

% The methods of \cite{Banks-Green, Kallosh-Rajaraman, Berenstein-Trnacanelli, Mazzucato-Vallilo} rely on studying higher-derivative corrections of Type II B supergravity or $\alpha'$-corrections in the AdS super-coset sigma model.

A trivial yet key observation that we will exploit is that rescaling an AdS $\times M$ geometry by a constant factor results in the same metric, but now with the AdS length re-scaled by the scaling factor. This statement is true both in the Poincare patch as well as in global AdS coordinates, and is independent of the dimensions of both the AdS as well as the compact geometry $M$. In both global and Poincare cases, this process involves a rescaling of the boundary coordinates. In Poincare AdS, we will often work with an arbitrary (but fixed) size boundary grid, so this re-scaling is not as interesting as it is in the global case where the cylinder comes with its own size, the AdS length scale.

For global AdS$_{d+1} \times S^q$, by introducing the notation
\bea
ds^2_{AdS_{d+1} \times S^q}(L)\equiv -\Big(1+\frac{R^2}{L^2}\Big)dt^2+\frac{dR^2}{\left(1+\frac{R^2}{L^2}\right)}+R^2d\Omega_{d-1}^2+L^2d\Omega_q^2
\eea
we can write this crucial fact in the form
\bea
\lambda^2\ ds^2_{AdS_{d+1} \times S^q}(L)= ds^2_{AdS_{d+1} \times S^q}(\lambda L).
%\left(-\Big(1+\frac{r^2}{L_0^2}\Big)dt^2+\frac{dr^2}{\left(1+\frac{r^2}{L_0^2}\right)}+r^2d\Omega_3^2+L_0^2d\Omega_5^2\right)=\left(-\Big(1+\frac{R^2}{L^2}\Big)dT^2+\frac{dR^2}{\left(1+\frac{R^2}{L^2}\right)}+R^2d\Omega_3^2+ L^2d\Omega_5^2\right)
\eea
This is easily checked by redefining the $R$ and $t$ coordinates. We will use analogous notation to describe analogous operations in other AdS geometries as well. % A similar calculation can also be done in the Poincare-type metric \eqref{NHmetric}, and one again sees that the effect of the scaling operation on the metric is to change the AdS length scale \eqref{LvsN} by a factor of $\lambda$.

As mentioned in the introduction, we wish to explore the idea that sub-AdS holography is controlled by a sub-stack $M$ of the $N$ total number of D-branes. We will motivate this further in the remaining sections, but for now the picture is simply that in a stack of $M$ D-branes the relevant length scale is the one controlled by the corresponding long string length. By our observations in the previous paragraph, it should be immediately clear that one can obtain the long string near-horizon metric \eqref{NHmetric} of a stack of $N$ D-branes by scaling the string scale metric (with $N=1$) by a factor of $N^{1/2}$. When we do this, the AdS length becomes \eqref{LvsN}.   Now, the claim that sub-AdS scales are controlled by sub-stacks of $M$ branes has a natural implementation -- instead of rescaling the metric by a factor $N^{1/2}$, we simply rescale by a factor of $M^{1/2}$ (or equivalently, rescale the string length $l_s$ by a factor of $M^{1/4}$). The result is simply the usual AdS$_5 \times S^5$ metric, but with an AdS length scale given by
\bea
R_0^4= 4 \pi g_s l_s^4 M.
\eea
We expect that the physics of states that are localized at sub-AdS scales $R_0$ in the original AdS$_5 \times S^5$ (with length scale $L$) can be understood by viewing them as states in the new AdS$_5 \times S^5$ with length-scale $R_0$. We will illustrate this philosophy throughout the paper. 

It is convenient to formulate these statements in terms of a central charge. The central charge of an AdS$_{d+1}$/CFT$_d$ system is (see e.g., \cite{dHoker})
\bea
\frac{c}{12} = \frac{c(L)}{12}\equiv \frac{A(L)}{16 \pi G_{d+1}(L)} \label{central}
\eea
where $A(L)=\Omega_{d-1} L^{d-1}$. This formula \eqref{central} reduces to the celebrated Brown-Henneaux \cite{Brown} central charge in AdS$_3$.  It should be noted that $G_{d+1}$ is the Newton's constant in $(d+1)$-dimensions, and we have emphasized with our notation that it depends on the volume of the compact dimension. We can (and often will) view the AdS$_{d+1}$ as arising from an AdS$_{d+1} \times S^q$ compactification with a higher dimensional $({d+q+1})$-dimensional Newton's constant. 

The reason we will be interested in the central charge is because it is a quantity that is naturally defined {\em at} the AdS length scale, and yet is naturally viewed as being constant at super-AdS scales. It is a measure of the number of gauge degrees of freedom accessible, which is a constant at super-AdS scales (and scales as eg., $N ^2$ in the AdS$_5 \times S^5$ case). The discussions in the previous paragraphs therefore naturally suggest that we associate a central charge to a sub-AdS scale via the central charge of the corresponding rescaled AdS geometry: 
\bea
\frac{c(R_0)}{12}= \frac{A(R_0)}{16 \pi G_{d+1}(R_0)}
\eea
Note that since the $(d+1)$-dimensional Newton's constant depends on the volume of the compact space, in terms of the the higher dimensional Newton's constant, this reads
\bea
\frac{c(R_0)}{12}=\frac{ \Omega_{d-1} \Omega_{q} R_0^{d+q-1}}{16 \pi G_{d+q+1}}=\Big(\frac{R_0}{L}\Big)^{d+q-1} \frac{c(L)}{12} \label{central-scaling}
\eea
For specific AdS geometries, the ratio of the AdS length scales will be related to the ratio of the sizes of the D-brane stacks. For AdS$_5 \times S^5$ for example, from our earlier discussion it follows that 
\bea
\frac{c(R_0)}{12} = \left(\frac{R_0}{L}\right)^{8} \ \frac{c(L)}{12} = \left(\frac{M}{N}\right)^{2} \ \frac{c(L)}{12}.
\eea

Closely related formulas (but with some important distinctions) have appeared in \cite{Verlinde}. They made various proposals for holographic screens in conformally related spacetimes as well as made connections to the long string in (twisted sectors of) AdS$_3$/CFT$_2$. In Appendix \ref{App-Verlinde} we show that our proposal avoids some of the unsatisfactory features of \cite{Verlinde} and therefore can be viewed as a strengthening of the original ideas in \cite{Verlinde}. 

\subsection{Manifestations of Long String Fractionation}

A key observation about long strings is that as the length of the long string increases, the energy of the minimal energy quantum in the system decreases. The seed of this idea goes back to the work of Das and Mathur \cite{Das} during the early days of the D1-D5 black hole \cite{StromingerVafa}. The picture is simple: the minimal energy quantum on a violin string of length $\ell$ goes inversely with $\ell$, see Maldacena and Susskind \cite{SusskindFat} for an  illuminating discussion.

What we have proposed in the previous subsection has the immediate consequence that the length of the (backreacted) long string increases as we go to longer distances in sub-AdS scales. Together with the statement above about the minimal energy quantum, this  means that the IR direction of the bulk in this regime corresponds to the IR scales of the microscopic description. This IR/IR duality in sub-AdS scales was argued in \cite{Verlinde} to be related to the negative specific heats of small black holes. Some of the consequences of IR/IR duality for Page curves of flat space black holes were discussed in \cite{Jude} and more evidence from flat space Ryu-Takayanagi prescriptions for such an IR/IR duality were noticed in \cite{Abir}.  

The minimal energy quantum $\epsilon_{dof}$ associated to a region of space can be obtained from the  physics of the largest black hole in that region. A measure of $\epsilon_{dof}$ is given by the energy/mass of such a black hole divided by the maximum number of degrees of freedom in that region. The latter is simply the area/entropy of the black hole. This means that the minimal energy quantum scales as 
\bea
\epsilon_{dof} \sim \frac{M}{\mathcal A}.
\eea 
It is easy to check \cite{Verlinde} for a horizon of size $R$, that $\epsilon_{dof}$  scales $\sim R/L^2$ for large AdS black holes, while it scales as $\sim 1/R$ for flat space (or small AdS) black holes. The former is consistent with the IR/UV duality expected at super-AdS scales, while the latter can be taken as a manifestation of IR/IR duality on sub-AdS scales.

It is plausible that the sign of the specific heat of a black holes is closely related to how $\epsilon_{dof}$ scales with the size of the horizon. To see this, note that the specific heat is given by $C=T dS/dT$. Since $T$ is always positive, it is clear that the sign is determined by $dS/dT$. Now, the area/entropy of a black hole is an increasing function of its radius quite universally, but the temperature need not be. For large AdS black holes the temperature is an increasing function of the horizon size, while for small or flat space black holes, it decreases with size. (We will give some explicit formulas later, but this qualitative feature will be enough for our purposes here.) Together with the discussion in the previous paragraph, this shows that the signs of $d \epsilon_{dof} / dR$ and $d T/dR$ (and therefore $C$) are correlated in the cases of interest to us.

These observations are somewhat heuristic, but they are also suggestive. We will find further evidence for them beyond those noticed in \cite{Verlinde, Jude, Abir} in later sections.

\subsection{Other Examples}\label{D1D5}

The discussion above is clearly general enough to incorporate large classes of AdS compactifications. Partly to demonstrate this, but also to comment on some specifics, we present the case of the D1-D5 system and M-theory/ABJM theory in this subsection.

\subsubsection{D1-D5 system}

The near horizon geometry in this case is given by \cite{Maldacena}:
\bea
ds^2 = l_s^2\left(\frac{U^2}{g_6\sqrt{N}}(-dt^2+dx^2)+g_6\sqrt{ N}\frac{dU^2}{U^2}+g_6\sqrt{N}d\Omega_5^2\right) \label{NHD1D5metric}
\eea
where $N =Q_1Q_5$ is the power of the symmetric product $(T^4)^N/S_N$ where we take the compact space to be $T^4$ for concreteness. The six-dimensional string coupling $g_6 \equiv 4 \pi^2 g_s l_s^2/\sqrt{V_4}$ along with $g_s$ and $r/l_s^2$ are held fixed as $l_s \rightarrow 0$ in the decoupling limit. The string length multiplying the 6d metric above will get rescaled under the long string operation -- this is natural because as we noted in the D3-brane system, the metric is measured in these units. The AdS length scale of the $M$-rescaled metric is 
\bea
R_0^4= g_6^2 \ l_s ^4 \ M, 
\eea
which corresponds to a sub-AdS$_3 \times S^3$ radius. When $M \rightarrow Q_1 Q_5$ we have the Maldacena formula for the AdS$_3$ length scale, $L^4 = g_6^2\ l_s^4 Q_1 Q_5$.

\subsubsection{ABJM, M-theory and the Fractionation of Branes vs Strings}

The situation with M2/M5-branes is interesting because in M-theory, there is no direct notion of string length. But the notion of fractionation \cite{Das, SusskindFat} is not restricted to fundamental strings, so we expect that a similar story may hold.

We will work with M2-branes, because the dual gauge theory in various limits is known \cite{BLG, ABJM}. The decoupling limit of M2-branes \cite{Maldacena} is $l_P \rightarrow 0$ while holding $U \equiv r^2/l_P^3$ fixed, where $l_P$ is the 11-dimensional Planck's constant. We will write the resulting metric as
\bea
ds^2= l_P^2\left(\frac{U^2}{(2^5 \pi^2 N k)^{2/3}}(-dt^2+d{\bf x}^2)+\frac{(2^5 \pi^2 N k)^{1/3}}{4}\frac{dU^2}{U^2}+(2^5 \pi^2 N k)^{1/3}d\Omega_{S^7/\IZ_k}^2\right) \label{M2metric}
\eea
The {\bf x} are the 2 transverse spatial dimensions of the M2-brane. The length scale is\footnote{One has to be a bit careful in identifying the charge with $N$. When $k >1$, there is a correction \cite{Hirano}, but this distinction will not affect our discussion.} 
\bea
L^6 = 2^5 l_P^6 \pi^2 N k \label{M2L}. \eea 
This is the sphere radius and the AdS length is half this, but we will refer to both loosely as the AdS length. Note that we have allowed the possibility of connecting to the IIA description via reduction on a circle fiber on the $S^7$, with the possibility that the circle is quotiented by a $\IZ_k$. When we take $k=1$, we have the near-horizon geometry of M2-branes in flat space, but when $k >1$ the original branes are at the tip of a cone and the compact geometry becomes a $\IZ_k$ quotient of a circle $S^1$ fibered over base $\IC P^3$. The latter is a Hopf fibration description of the $S^7$.  When $N \rightarrow \infty$ with fixed $k$, we are in the M-theory phase of ABJM gauge theory. 

The degrees of freedom describing the interaction between M2-branes in an M2-brane stack, are not fully known. This is unlike in the case of D-branes in string theory where we know that the interaction is mediated by open strings, in the perturbative limit. It has been suggested in \cite{Shahin}, based on the 3-algebra description associated to the Bagger-Lambert-Gustavsson path to M2-brane world-volume theory \cite{BLG}, that the relevant degrees of freedom are those of open M2-branes stretched between M2-branes in the stack. Either way, it is natural in the M-theory description \eqref{M2L} that the correct re-scaling associated to sub-AdS scales is 
\bea
R^6 = 2^5 l_P^6 \pi^2 M k \label{M2R}
\eea 
which treats the Planck length instead of the string length as the basic length scale. We will see later that this scaling correctly reproduces the entropy scaling of small black holes localized on the $S^7$ of AdS$_4 \times S^7$. 

What makes ABJM theory interesting from our perspective is that it also has a stringy limit.  The metric \eqref{M2metric} in the form we have written can be connected to the IIA description where there is an explicit string available. The reduction is along the circle fibered over the $\IC P^3$. We write the 11d metric as\footnote{We are being explicit to keep track of the $l_P$ and the $l_s$ dependencies, which are often suppressed \cite{ABJM}. We essentially follow the conventions of \cite{Headrick} for the reduction.} 
\bea
\frac{1}{l_P^2}  {ds^2_M}= \tilde L^2 G_{\mu\nu}^0 dx^\mu dx^\nu+\frac{\tilde L^2}{k^2}(d \phi + k\omega)^2, \ \ \phi \sim \phi+2 \pi.
\eea
where $G^0_{\mu\nu}$ stands for the AdS$_4$ together with the base $\IC P^3$.  We have used tildes to denote the 11d dimensionless AdS length, $\tilde L \equiv L/l_P$.  After the reduction this leads to the IIA form (in string frame)
\bea
\frac{1}{l_s^2} ds^2_{IIA} = \frac{\tilde L^3}{k}  G^0_{\mu\nu}dx^\mu dx^\nu, \ \ e^{\Phi} = g_s = \Big( \frac{\tilde L}{k}\Big)^{3/2}
\eea
Note that the usual M-theory reduction formula $l_P = g_s^{1/3} l_s$ leads to  $l_P=(\tilde L/k)^{1/2} l_s$. It follows from the above expressions that the dimensionless IIA AdS radius is given by $\tilde L_{IIA}^2 \equiv \frac{L_{IIA}^2}{l_s^2}= \frac{\tilde L^3}{k}$. At this stage, we can use \eqref{M2L} to show that 
\bea
 L^2_{IIA} = \frac{\tilde L^3}{k} l_s^2 = 2^{5/2} \pi \sqrt{\frac{N}{k}} l_s^2
 \eea
This (perturbative) IIA string arises when $\lambda \equiv \sqrt{N/k}$ is held fixed and $N \rightarrow \infty$. This is IIA phase of ABJM gauge theory. The AdS length-scale and its dependence on $N$ are both different from \eqref{M2L}. The advantage compared to the M-theory phase is that we no longer have to guess what the inter-brane degrees of freedom are, they are open strings. This leads immediately to the long-string re-scaled result at sub-AdS length scales $R_{IIA}$:
\bea
 R^2_{IIA} = 2^{5/2} \pi \sqrt{\frac{M}{k}} l_s^2.
\eea
This scaling can correctly reproduce the thermodynamics of the small black hole localized on the $\IC P^3$ of AdS$_4 \times \IC P^3$. This follows as a trivial variation of the results in the next sub-section. The scaling of the AdS-length is $R_{IIA}\sim M^{1/4}$ which is the same as what we found in AdS$_5 \times S^5$. 

It will be very instructive to understand the dynamical origins of the M-brane fractionated phases vs long string phases of ABJM theory.

\subsection{Small Black Holes}

Our long string re-scaling suggests that we should associate a rescaled AdS $\times$ S geometry to the sub-AdS scale, with the new AdS length equal to that scale. It is natural to try and understand small black holes of the original AdS in terms of {\em large} black holes of the rescaled AdS. Alternatively, we can use the physics of small black holes as a touchstone for our proposal, by viewing them as large black holes of this re-scaled AdS $\times$ S and seeing whether any of the expectations about them are correctly reproduced. The possibility of viewing small black holes as large black holes in a ``smaller" AdS has appeared before \cite{Hanada}. The idea that a small black hole is a deconfined configuration of a sub-matrix or a sub-stack of D-branes, is a recurring theme in this paper.

We start by reviewing a few basic formulas for large black holes in AdS$_5 \times S^5$. The metric takes the form
\bea
ds^2 = -\Big(1+\frac{r^2}{L^2}-\frac{\mu}{r^2}\Big) dt^2 + \frac{dr^2}{\Big(1+\frac{r^2}{L^2}-\frac{\mu}{r^2}\Big)} + r^2 d\Omega_3^2 + L^2 d\Omega_5^2 
\eea
The mass parameter $\mu$ is related to the horizon radius $r_+$ via
\bea
\mu=\frac{r_+^4}{L^2} + r_+^2,
\eea
and the Hawking temperature is
\bea
T=\frac{2r_+^2+L^2}{2 \pi L^2 r_+}.
\eea
The entropy is given by 
\bea
S= \frac{ \pi^2 r_+^3}{2 G_5}=\frac{2 \pi^2 r_+^3 \times \pi^3 L^5}{4 G_{10}},
\eea
where in the second equality, we have written the expression in a form where the 10-dimensional origins are transparent.  We can reproduce the correct scaling behavior of the entropy of the small black hole as well as the negativity of its specific heat by viewing it as a large AdS black hole in a rescaled AdS $\times$ S geometry with AdS length scale equal to the horizon radius. This is easily seen, by replacing the AdS length scale $L$ with $r_+$ in the above expressions. The entropy now scales as $ S \sim \frac{r_+^8}{G_{10}}$, 
which is the expected correct scaling \cite{Hanada} and matches that of the flat space 10 dimensional black hole.  The expressions for mass parameter and temperature become $\mu \sim r_+^2$ and $T \sim \frac{1}{r_+}$. This leads to $\mu \sim \frac{1}{T^2}$ and as a result, the specific heat goes as $S = \frac{d \mu}{d T} \sim -\frac{1}{T^3}$ which is manifestly negative.  

The structure of these results generalize straightforwardly to other Freund-Rubin type AdS compactifications, so we will not dwell on the details. For example, it should be clear that the large AdS black hole in an AdS$_p \times M^q$ geometry has an entropy that scales as $\frac{r_+^{p-2} \times L^q}{G_{p+q-2}}$, while the small black hole localized on the $M^q$ has an entropy that goes as $\frac{r_+^{p+q-2}}{G_{p+q-2}}$.

When a black hole is localized on the $S^5$, we expect the solution to break the $SO(6)$ isometry and preserve only an $SO(5)$. Therefore it is natural to think that the re-scaled AdS$_5 \times S^5$ which controls its physics should also be viewed as being localized on the $S^5$. Further evidence and implications for this will be presented in the next section, by investigating the Coulomb branch at finite temperature \cite{Kraus}.

\section{AdS/CFT on the Coulomb Branch}

In ${\mathcal  N}=4$ $SU(N)$ SYM theory, the Coulomb branch is defined by non-vanishing VEVs for the scalar fields $\Phi_i$ ($i=1,..,6$) that satisfy $[\Phi_i, \Phi_j]=0$. The moduli space can be characterized by 6$(N-1)$ eigenvalues arising from the 6 diagonalized traceless matrices $\Phi_i$. At a generic point on the Coulomb branch, the gauge symmetry is broken to $U(1)^{N-1}$ and the low energy effective action below the scale of the VEVs can in principle be found by integrating out the massive off-diagonal degrees of freedom. The Coulomb branch breaks conformal invariance, but preserves the $\mathcal{N}=4$ supersymmetry of the theory and the coupling is not renormalized (which is related to the constancy of the dilaton in D3-brane  supergravity solution).

The gravitational description of the Coulomb branch can be done  via separating the D3-branes in the 6 transverse directions. These multi-center SUGRA solutions are completely characterized by harmonic functions of the transverse coordinates. We will be interested in certain classes of such solutions, which we will argue are closely related to localized black holes on the $S^5$. We write the D3-brane metric as
 \bea
ds^2= H^{-1/2}(r,\Omega_5)(-dt^2+\sum_{i=1}^{3}dx_i^2) +H^{1/2}(r,\Omega_5) (dr^2+ r^2 d\Omega_5^2)
\eea
where the SUGRA equations are solved for any harmonic function $H$ of the transverse coordinates, which we have denoted by $(r, \Omega_5)$ in terms of a radial coordinate and the 5-sphere. 

\subsection{Multi-Center Solutions}

\begin{figure}[H]
    \centering
    \includegraphics[width=0.8\linewidth]{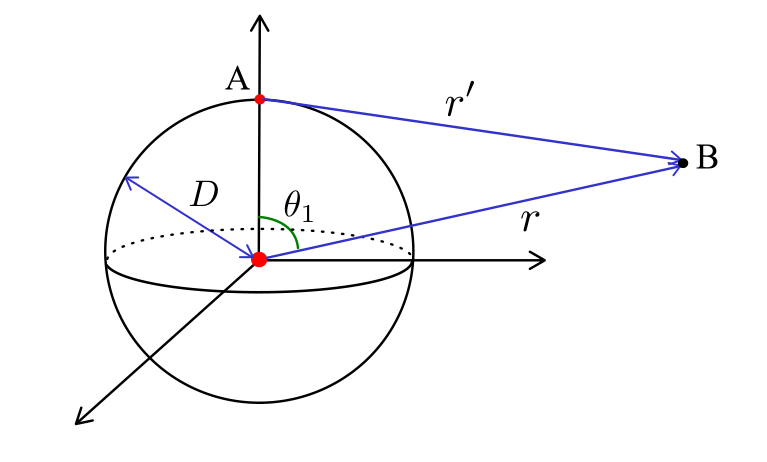}
    \caption{A Coulomb branch configuration with two D-brane stacks marked as red dots. The sphere stands for an $S^5$. The $r$ direction together with sphere denotes the six transverse directions, the longitudinal brane directions are suppressed. }
    \label{Coulomb-Stacks}
\end{figure}

Let us consider a harmonic function that is generated by two sources as depicted in Figure \ref{Coulomb-Stacks}. 
One of the sources is localized at the origin, and the other is separated by a distance $D$ from the origin in the transverse directions (denoted by point $A$). Without loss of generality, we can place it at the North pole, which is specified by $\theta_1=0$ upon introducing a suitable polar angle on the 5-sphere. The total harmonic function at a generic point with coordinates $(r, \theta_1, ...)$ in the transverse space is
\bea
H\equiv 1+H_0+H_A =1+ \frac{L^4-R^4}{r^4} + \frac{R^4}{(r^2+D^2-2D r \cos \theta_1)^2}. 
\eea
The form should be clear, the piece that requires some explanation is the strength of the sources. The idea here is that we have split the D-branes into two stacks. The one at $A$ contains $M$ D-branes, while the one at the origin contains the rest, $N-M$. The length scale $L$ is controlled by \eqref{LvsN} while $R$ is defined via
\bea
R^4 = 4 \pi g_s l_s^4 M. \label{RvsM}
\eea
We will take $R < L$ and $ \sqrt{\alpha'} \ll D \ll R, L$. When $r$ is larger than all other length scales in the problem, it is clear that the solution reduces to the conventional D3-brane supergravity solution with total strength $L$. The near horizon limit on the other hand involves multiple scales. In the regime where $D \ll r \ll L, R$ we can drop the 1 in the warp factor, but there is a useful notion of the total stack so that the geometry is an AdS throat with length scale $L$. But if we consider shorter distances and zoom in close to either of the stacks, there are two separate sub-AdS throats with length scales controlled by $(L^4-R^4)^{1/4}$ and $R$. What we have described is therefore a specific point on the Coulomb branch which is asymptotically AdS$_5 \times S^5$ with length scale $L$, but differs from it in the interior.

\subsection{Localized Black Holes on the $S^5$}

The utility of the above construction is that it has a natural connection to small black holes localized on the $S^5$ of the near-horizon Poincare AdS$_5 \times S^5$ geometry. Small black holes are black holes small enough to look Schwarzschild-like near the horizon (in ten dimensions). They are localized both on AdS$_5$ and $S^5$, which is distinct from large AdS black holes, whose horizon is localized in AdS$_5$ but covers the entire $S^5$. Small black holes cannot be dominant phases thermodynamically, because they can evaporate away. 

Our discussion of small black holes in a previous section was in the context of global AdS$_5$. Below the Hawking-Page temperature, this system is dominated by a hot gas of gravitons into which the small black hole evaporates. The phase transition between the gas of gravitons and the large-AdS black hole is dual to the deconfinement transition in the gauge theory on a compact space. In Poincare AdS$_5$ on the other hand, at a given temperature, the system has a finite energy {\em density}. At fixed CFT temperature (no matter how small), the dominant phase is the planar AdS black hole. The small black hole localized on the $S^5$ with some horizon temperature\footnote{This cannot be attributed as the CFT temperature, because the energy density is zero after evaporation because of the non-compact brane directions.}, evaporates to a configuration with arbitrarily low temperature and energy density, that is indistinguishable from the vacuum planar AdS geometry.  Note that while not thermodynamically stable, we expect these black holes to be long-lived metastable configurations unstable only to Hawking decay.

\begin{figure}[H]
    \centering
    \includegraphics[width=0.5\linewidth]{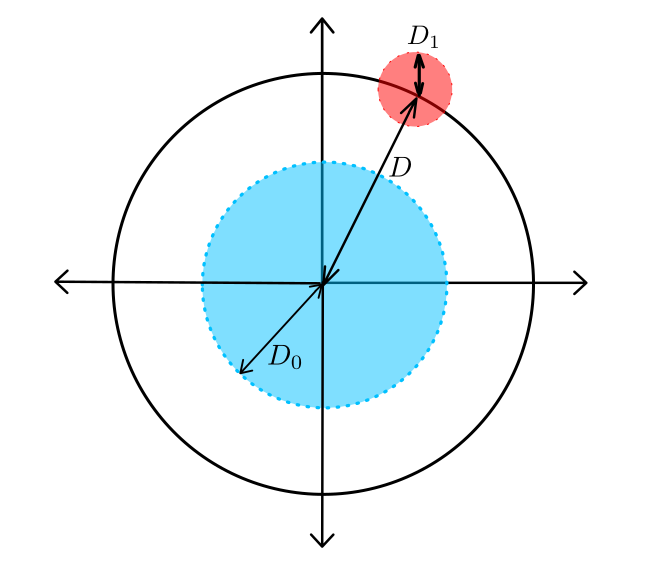}
    \caption{A Coulomb branch configuration that can heat up to a localized black hole on the $S^5$ of planar AdS$_5 \times S^5$. We have denoted the sphere by a circle, the radial direction is the transverse radial direction, and the longitudinal brane directions come out of the plane of the figure.}
    \label{Coulomb-Stacks2}
\end{figure}

The Coulomb branch solution we constructed earlier in this section, when heated up, can result in a localized Poincare AdS$_5$ black hole of this type. The key idea is that {\em a localized black hole arises from heating up a sub-stack of branes to a temperature that is comparable to the length-scale $R$ produced by the D-brane sources in that sub-stack.} This is natural from the perspective that small black holes are the result of sub-matrix deconfinement \cite{Berenstein, Hanada}, because $R$ is the length-scale \eqref{RvsM} associated to the $ M \times M$ sub-matrix in the gauge theory description. This is also what we needed for the area-scaling of the small black hole entropy to work out correctly. 

Note that once we heat up the branes as proposed above, there is a Gregory-Laflamme instability in the longitudinal spatial directions of the brane. This is crucial for ensuring that the end result is a black {\em hole} localized in all directions rather than a black brane localized only on the $S^5$ and not AdS$_5$. It is natural that the scale of the clumps formed at the end of the Gregory-Laflamme instability is controlled by the length-scale $R$. Note also that if we were to zoom in on a localized stack, and then heat it up, we would instead find the usual planar black brane which covers the entire $S^5$ (which crucially, does not have the Gregory-Laflamme instability\footnote{We thank Roberto Emparan for an explanation on the stability properties of planar AdS horizons. We give a qualitative discussion of the role of Gregory-Laflamme instability, the near horizon limit of the total stack, and what makes the localized stack different, in an Appendix.}).

Let us now discuss this in more detail, and identify some conditions on the relevant scales in the problem. As mentioned earlier, in order for the multi-center solution to have an interpretation as an asymptotically AdS solution with length scale $L$, it must be the case that there is a hierarchy between $D$ and $L, R$. This is necessary for a regime of $r$ such that $D \ll r \ll L,R$ exists, where one can take a near-horizon limit of the total stack. We wish to view the small black hole localized on the $S^5$ as a finite temperature configuration in which the $R$-stack sub-matrix has deconfined. As suggested above, this means that the Hawking temperature of the small black hole is determined by the scale $R$, i.e., $T_H \sim \frac{1}{R}$. But we also need that this temperature is not so high, that the strings stretched {\em between} the stacks (which require energies of the order $\sim \frac{D}{\alpha'}$) are excited. They would correspond to off-diagonal degrees of freedom of the matrix, and exciting them would be tantamount to deconfining a bigger (sub-)matrix. In other words, we have the conditions
\bea
\frac{1}{L} \lesssim \frac{1}{R} \sim T_H \ll \frac{D}{\alpha'}, \ \ \ {\rm and} \ \ \ D \ll L,R. \label{LReq}
\eea
Note that the temperature of the localized black hole is higher than the would-be Hawking temperature of the whole stack (because $R$ is smaller than $L$). One might worry that this would lead to deconfinement of the whole stack. The key point however is that it is a localized temperature in the bulk. It is inequality \eqref{LReq} that arranges this -- what it shows is that the two brane stacks are separated by a distance $D$ and the localization of the temperature arises because in order to excite the whole system, we need to excite open strings stretched over $D$. This picture has a natural understanding in the fully backreacted gravity picture as well, which we will discuss at the end of the next subsection.

The discussions above considered the situation in Figure \ref{Coulomb-Stacks} which started with the zero-temperature configuration where the two stacks of branes were localized as delta functions. We can relax this and consider the possibility that the brane stacks are somewhat delocalized. This is shown in Figure \ref{Coulomb-Stacks2}. The sizes of the stacks are what determine the open string excitation energy required to excite the off-diagonal modes within them\footnote{This should not be confused with the number of D-branes within the stack, which is what controls the scales $L$ and $R$.}. In this slightly more general setting, we will also need to impose that 
\bea
\frac{D_1}{\alpha'} \lesssim T_H \ll \frac{D_0}{\alpha'}.
\eea
The first condition demands that the delocalization of the branes in the small black hole stack should not exceed the excitation energy available from the temperature of the black hole. The second condition enforces that the second stack is not deconfined. 

Note that the {\em asymptotically} AdS geometry that characterizes the Coulomb branch, with a stack of multi-center branes deep in the bulk, was crucial for this discussion. If instead we zoom in on any one of the delta-function stacks and then turn on the temperature, we will end up with the usual planar AdS black hole associated to that stack. Irrespective of the magnitude of the temperature, we will now be in the fully deconfined phase because the horizon is planar. We will have more to say about analogous situations with more general Dp-branes in the next section.

\subsection{Localizing Re-scaled AdS$_5 \times S^5$ on AdS$_5 \times S^5$: The Global Case}

Our arguments so far suggest that the small black hole is naturally viewed as a ``heated up" re-scaled AdS$_5 \times S^5$ geometry localized on the original AdS$_5 \times S^5$. Our discussion was in the setting of the Poincare AdS geometry, which is the one most directly accessible in the near-horizon limit. As we briefly discussed, in Poincare AdS, a black hole localized on the $S^5$ cannot be viewed as a (perhaps unstable) CFT state at fixed non-zero temperature. The reason is that when it evaporates, it does so to an infinite phase space and therefore the final temperature is effectively zero. 

The situation is different when we consider a global AdS$_5 \times S^5$ black hole. Small black holes of global AdS$_5 \times S^5$ can evaporate to a gas of hot gravitons in the bulk. Because of the box nature of AdS, we do not expect the final temperature to be zero. While this does not change the status that the small black hole is still not a thermodynamically stable phase, it does behoove us to provide a separate discussion. A key underlying point is that it is the planar and not the global AdS geometry that arises from a near-horizon decoupling limit of the original flat space branes. Therefore our Coulomb-branch motivated approach of the last section will not work. Instead, we are looking for a geometry that has a {\em global} AdS$_5 \times S^5$ geometry with a length scale $R$, localized at a point of a ``bigger" global AdS$_5 \times S^5$ with length scale $L$. 

Our goal in this subsection will simply be to demonstrate that ansatzes for such geometries are easy enough to write down in the setting of the small AdS$_5 \times S^5$ black hole numerically constructed by Dias, Santos and Way \cite{DSW}. We wish to write down an explicit metric that can be viewed as a {\em rescaled} global AdS$_5 \times S^5$ living inside a ``bigger" global AdS$_5 \times S^5$. Note in particular that simply rescaling the AdS length will change the asymptotic structure as well, so this is not what we want -- we want a rescaled AdS that can be viewed as being localized on the $S^5$, while the asymptotic behavior is still that of the bigger Ad$S_5$. Despite the heuristic discussions in earlier papers \cite{Hanada,Verlinde} an explicit metric which has these properties was never written down to the best of our knowledge.

In \cite{DSW, DSW2}, small black holes localized on global AdS$_5 \times S^5$ were constructed by starting with a suitable ansatz, and then solving for the arbitrary functions in the ansatz under the equations of motion of type IIB supergravity. We will simply notice that elementary choices of this ansatz allows us to write down localized global AdS$_5 \times S^5$ spacetimes. The ansatz used in \cite{DSW} is somewhat complicated, we will present a restricted version of it here because that will be sufficient to illustrate our point. Let us consider the metric
\bea
ds^2= -M dt^2 + L^2  f(\rho, \xi) \left[d\rho^2+ \rho^2\Big( \frac{d\xi^2}{2-\xi^2} + G_1 \xi^2(2-\xi^2) d\Omega_3^2+G_2 (1-\xi^2)^2 d\Omega_4^2\Big)\right]. \label{LocalAdS}
\eea
Here $L$ is the overall AdS length scale. We will fix 
\bea
M \equiv \cosh^2(\rho \xi\sqrt{1-\xi^2}), \  \ G_1 \equiv \Big({\rm sinc} (i \rho \xi \sqrt{2-\xi^2} \ )  \Big)^2, \ \ G_2 \equiv \Big({\rm sinc}\  \rho \sqrt{1-\xi^2})  \Big)^2 
\eea
in all of our discussions. Here ${\rm sinc}\ x \equiv \frac{\sin x}{x}$.  If we also set $f \equiv 1$, this metric is in fact\footnote{The philosophy behind writing the AdS$_5 \times S^5$ metric in this way can be understood in a simpler context. Consider a metric on $S^1 \times S^1$ given by $ds^2=dx^2+dy^2$, with $x\sim x+2 \pi$ and $y \sim y+2 \pi$. To manifest structures that are localized at points on each of the circles, it is useful to introduce a locally polar coordinate system around that point, $ds^2=d\rho^2+\rho^2 d\phi^2$. A similar statements applies to $ds^2=ds^2_{AdS_5}+ds^2_{S^5}$. By a suitable choice of coordinates \cite{DSW}, the point around which this ``polar coordinates" is written can be taken to the origin of AdS$_5$ and the pole of $S^5$.} exactly AdS$_5 \times S^5$ with length scale $L$. The advantage of this form is that it provides a type of interpolation between flat 10 dimensional space and AdS$_5 \times S^5$. To appreciate this, set the functions $M, G_1$ and $G_2$ as well as $f$ to 1. We find that the metric is simply 10 dimensional Minkowski space. This can be understood as a $\rho \rightarrow 0$ limit.  $M, G_1, G_2$ trivialize in that limit, and we find the $S^8$-sphere of 10-dimensional flat space as an $S^3 \times S^4$ fibered over a segment parametrized by $\xi$.

The discussion in \cite{DSW} involved two more fixed functions which captured the black hole horizon radius. When the horizon radius is set to zero, these functions reduce to unity, so we have not included them here. There are also more unfixed functions in \cite{DSW}. We have set some of them to zero, and the rest equal to each other. What remains is a single unfixed function of $\rho$ and $\xi$, which we have called $f(\rho, \xi)$ above. Our goal here is to identify a minimal setting where we can embed a localized AdS$_5 \times S^5$ in an intuitive way, so the extra functions will not be needed.

The ansatz and the discussion above about the flat space limit immediately suggests what we should do to obtain a localized AdS$_5 \times S^5$. We need to choose a function $f(\rho ,\xi)$ so that at large $\rho$ it tends to 1 and the geometry reduces to AdS$_5 \times S^5$ with length scale $L$, while at small $\rho$ we want $f(\rho,\xi)$ to tend to $\frac{R^2}{L^2}$ so that the effect on the metric \eqref{LocalAdS} is simply that we have AdS$_5 \times S^5$ with length-scale $R$. A simple example of a function that does this, is
\bea
f(\rho,\xi)=\frac{R^2}{L^2}+\Big(1-\frac{R^2}{L^2}\Big)\tanh \Big(\frac{\rho}{\rho_0}\Big).
\eea
Of course, the point about this function is only that it is quite simple and explicit. We have taken it to be independent of $\xi$ for simplicity, but any function of $\rho$ and $\xi$ that has the asymptotic behaviors at large and small $\rho$ that we outlined above will do the job. It is also worth pointing out that we have taken a simple ansatz to illustrate the structure here, one can also work with the more general setting of \cite{DSW} which had more undetermined functions $f_{1,3,5}$ and $\tilde f_{2,4,6}$ (the functions $H_1$ and $H_2$ there, we can set to 1, since we are not discussing black holes directly). In fact, it will be interesting to look for a solution of the Type IIB supergravity system with an ansatz for the 5-form flux also turned on, which admits a localized AdS$_5 \times S^5$ solution of this form. This will require turning on D3-charge sources for the flux, since the flux as captured by the AdS length scale is changing with $\rho$ (and presumably $\xi$). This will involve solving PDEs numerically, and we will not undertake it. The possibility of a solution with the D3-charge smeared over the angles (in particular $\xi$) so that we can work with ODEs in $\rho$, is currently being investigated \cite{ToAppear}.

Our comments on heating up the localized stack in the previous subsection, dovetail nicely with the backreacted global case here, when we consider black holes. Because we are working with global AdS, we have a fairly complete understanding of this system via the usual Hawking-Page discussion. We see why the full system does not deconfine despite the higher temperature of the localized black hole -- the answer is that there are two possible black holes of different radii for any given Hawking temperature. Only the large black hole corresponds to the fully deconfined phase, because the temperature of the smaller black hole does not have a direct CFT interpretation (it is a local temperature in the bulk). This is because the small black hole is not a thermodynamically stable phase -- for example, as it evaporates, its Hawking temperature will keep increasing as is the case for flat space black holes. This discussion is in the context of the original Hawking-Page discussion, but the general ideas are expected to carry through to the explicitly (numerically) constructed localized AdS$5 \times S^5$ black hole of \cite{DSW}.

\section{Dp-brane Theories with 16 Supercharges}

In this section, we make some comments about (non-conformal) Dp-brane theories with 16 supercharges \cite{Itzhaki, Magoo}. We will restrict our attention to the $p < 5$ cases -- we wish to avoid the usual subtleties \cite{LST} associated to the 5-(and above-)branes here, but we hope to return to them in the future. 

The decoupled Dp-brane gauge theories have a dual description in terms of supergravity solutions in some regimes of parameters \cite{Itzhaki}. A crucial feature of these solutions is that they emerge naturally as conformally {\em Poincare} AdS geometries in the decoupling limit. We will argue below that these Poincare geometries are naturally associated to deconfined phases of the gauge theory, in a sense that we will clarify. This means that all the matrix degrees of freedom are excited, and therefore we are {\em not} in a regime where sub-matrix and long string effects will be important. We will show that this ties in precisely with our previous expectations about IR/UV vs IR/IR correspondence, specific heats of black holes, and the holographic bound (for Dp-branes). Even though not directly about long strings, we believe this ties in nicely with our arguments in a setting beyond standard AdS/CFT.

\subsection{The IR/UV Correspondence}

We start by writing the near-horizon Dp-brane metric in the familiar notation in the string frame \cite{Itzhaki}:
\bea
ds^2=\alpha'\left( \frac{U^{\frac{7-p}{2}}}{\sqrt{c_p g^2_{YM}N}}(-dt^2+dx_p^2)+\frac{\sqrt{c_p g^2_{YM}N}}{U^{\frac{7-p}{2}}}dU^2 +\sqrt{c_p g^2_{YM}N} U^{\frac{p-3}{2}}\right) 
\eea
with the dilaton
\bea
e^\phi = g^2_{YM}\ (g^2_{YM} N U^{p-7})^\frac{3-p}{4}.
\eea
Throughout this section, we will omit keeping track of overall numerical factors that do not affect the physics. The equalities should be understood as equalities up to such constants. 

The metric above is known to be conformal to Poincare AdS $\times  S$, as we illustrate in a few different coordinates in Appendix \ref{DpCoord}. It will be interesting to consider suitable global forms of these conformally AdS $\times S$ geometries, but they seem to have not been investigated. It would be interesting to investigate sub-matrix/long string effects in such geometries.

A basic observation \cite {Itzhaki} is that increasing $U = r/\alpha'$ is associated naturally to higher energies in the gauge theory. This can be motivated by viewing $U$ as a Higgs VEV or by viewing it in terms of extended open strings between D-branes. Either way, this implies that the increasing $U$-direction corresponds to the UV of the dual super-Yang-Mills  theory -- a statement independent of $p$. 

We will now argue that the correspondence between supergravity and gauge theory is an IR/UV correspondence. That is, we want to show that large $U$ corresponds to the IR of supergravity, irrespective of $p$. There are two natural ways to define IR in the bulk. The first is to consider the (radial) geodesic distance from $U=0$ to some finite $U=U_0$. It is easy to check that this is an increasing function of $U_0$ (this is true in both string frame and Einstein frame\footnote{In Einstein frame the relevant radial invariant length is given by $\Delta R_E \sim \int_0^{U_0}U^{\frac{(p-7)(p+1)}{16}}dU$.}), which means that larger $U$ implies longer radial distances in the bulk. A second notion of IR is to consider the invariant grid size of the transverse spatial dimensions $\Delta x_p$ at the location $U=U_0$. If this transverse size increases with $U_0$, then again larger $U$ can be viewed as the IR of the bulk. In Einstein frame, it is easy to check that
\bea
|| \Delta x_p||^2 \sim U_0^\frac{(7-p)(3-p)}{8}U_0^{\frac{7-p}{2}} |\Delta x_p|^2 \sim U_0^\frac{(7-p)^2}{8}|\Delta x_p|^2.
\eea 
This is clearly an increasing function of $U_0$. So both transverse and radial notions of ``size" increase with $U$ in the bulk, indicating that the IR of the bulk supergravity is in the direction of larger $U$. Note that these statements are again independent of $p$. 

Together, these observations indicate that for any value of $p$ the supergravity-gauge theory duality is an IR/UV duality. This is reminiscent of super-AdS scales in global AdS, and we will see that the physics is in many ways analogous. Let us emphasize that we are always working with the geometries obtained directly from the decoupling limit here, and therefore these geometries are all (conformally) Poincare AdS.

Even though the UV-direction of the gauge theory is in the direction of larger $U$, and this is independent of $p$, it is important to note that the UV-{\em behavior} of the gauge theory depends crucially on $p$. For $p < 3$ the gauge theory is UV-free, for $p > 3$ it is IR-free and for $p=3$ the coupling does not run. Interestingly, the supergravity description reflects this. The curvature scalar of the gravitational description becomes large at large $U$ for $p < 3$ while it becomes small at large $U$ for $p >3$ \cite{Itzhaki}:
\bea
\alpha' R \sim U^\frac{3-p}{2}.
\eea
Here $R$ stands for the Ricci scalar. For $p=3$, the curvature stays constant. These facts are directly related to the behavior of the dimensionless gauge coupling which goes as (see Appendix \ref{DpCoord}) $g^2_{eff} \sim g^2_{YM} N U^{p-3}$. 

\subsubsection{Holographic Bound in Dp-brane Theories}

In this section, we will compute the holographic bound in these theories, in analogy with the work of Susskind and Witten for AdS/CFT \cite{SusskindWitten}.  We loosely follow the discussion in \cite{Trivedi}. The result will be useful in the next subsection.

We will work with the string frame metric presented in \eqref{MagooDp}. To translate to Einstein frame, we will also need the dilaton presented earlier. The holographic bound on the screen at radius $z=z_0$ is given by\footnote{We will also use the notation $z_0$ to denote horizon radii of black branes later in this section, but this should not cause any confusion.}
\bea
N_{dof}  = \frac{A}{G_{10}} = \frac{A}{g_s^2 \alpha'^4} = \frac{A}{g_{YM}^4 \alpha'^4}\alpha'^{p-3}
\eea
In string frame this leads to 
\bea \label{HoloBoundString}
N_{dof} = \alpha'^{p-3} \frac{N^\frac{4}{5-p}}{(g^2_{YM})^{\frac{2(3-p)}{5-p}}} \frac{L_0^p}{z_0^\frac{9p-p^2-12}{5-p}}.
\eea
Here $L_0$ is the size of a spatial $x^p$-coordinate grid on the brane (not to be confused with any AdS length scale). When $p=3$ we see that this reduces to $N_{dof} = N^2 L_0^3/z^3$ which is the well-known result \cite{SusskindWitten, Trivedi} in AdS/CFT. In Einstein frame, a similar calculation leads to 
\bea
N_{dof}^E = \alpha'^{p-3} (g^2_{YM})^\frac{3p-13}{5-p} \ N^\frac{7-p}{5-p} \frac{1}{z_0^\frac{9-p}{5-p}}.
\eea
We will utilize these results in the next subsection.

\subsection{Black Branes and Full Deconfinement}

In standard AdS/CFT, the physics of global and Poincare AdS geometries have sharp distinctions. The former corresponds to a CFT on the cylinder, and therefore naturally comes with a scale which is naturally the AdS scale $L$. The natural deconfinement condition is $T \gtrsim 1/L$, which corresponds to large AdS black holes \cite{WittenBH}. For Poincare (or planar) AdS black holes, the length scale is infinite and therefore the deconfinement condition is simply $T >0$. In other words, all Poincare AdS black holes are in the deconfined phase, irrespective of how low the temperature is. Note also that the geometries that arise from the decoupling/near-horizon limit are naturally in the Poincare form. 

Geometrically, the picture here is that the horizon is compact in a global AdS geometry. So there is a natural sense in which the scale of the temperature can be bigger or smaller than the (inverse) ``size" of the horizon. The former corresponds to the deconfined phase. In the planar black hole on the other hand, the horizon is non-compact and therefore the temperature is always higher than the horizon ``size". The above argument suggests that it is natural for the {\em conformally} Poincare AdS brane geometries of the previous section to also be viewed as deconfined phases,  when they are placed at (arbitrary) finite temperature\footnote{We will assume that we are working in regimes where the 10d supergravity description is valid.}. The 
reason is that the horizon here is again, non-compact. The  planar black brane geometry in suitable coordinates is written for general $p$ in \eqref{BlBrane}.

We now present three related arguments inspired by our AdS discussions in the previous sections, that augment (and provide further evidence for) the above claims. Firstly, we note that in a fully deconfined phase of the gauge theory, the physics is associated to the largest matrix, and therefore the strings are always maximally long. This means that there is no long-string rescaling operation available, which in turn means that as we go to the IR of the bulk, the excitation energies cannot decrease. In other words, the duality cannot be an IR/IR duality. This is consistent with what we found in the previous subsection -- in the conformally Poincare geometry, the duality is an IR/UV correspondence. 

From our previous discussions, we expect that an IR/IR correspondence implies negative specific heat for black holes as well as an $\epsilon_{dof}$ that decreases with radius. An IR/UV correspondence on the other hand implies positive specific heat and an $\epsilon_{dof}$ that increases towards the IR of the bulk. We will compute both these quantities in the remainder of this subsection for planar Dp-brane black holes, and show that the result is again consistent with an IR/UV duality.   

\subsubsection{Specific Heat}

The overall conformal factor in the metric is regular at the horizon, so it can be absorbed into a rescaling of the near-horizon radial coordinate, and does not play a role in the determination of the temperature. The metric is of the form presented in  \eqref{BlBrane}, but we will do the calculation in the $z$-coordinate. The relevant part (modulo the precise conformal factor) is \eqref{BlackPart}. Demanding absence of conical defect gives
\bea \label{T_H}
T = \frac{2(7-p)}{5-p} \frac{1}{4 \pi z_0}
\eea
which matches with the planar black hole temperature of AdS$_5$ when $p=3$. Note that the $z$- and $U$-coordinates both measure the transverse brane coordinates in the same units -- i.e., there is no re-scaling of $t$ (or $x_p$) in going from $U$ to $z$. This is the normalization we use when computing the temperature here. If we work with $\tilde z$ or $\hat z$ instead, we will have to re-scale the transverse coordinates with a suitable power of $g^2_{YM}N$.  

To compute the specific heat, we first compute the entropy $S_E$ of the 10-dimensional Einstein frame solution and then apply $C=T\frac{dS_E}{dT}$. We find
\bea
S_E = \alpha'^{p-3}\ (g^2_{YM})^\frac{3p-13}{5-p}\ N^\frac{7-p}{5-p}\ z_0^\frac{p-9}{5-p},
\eea
which upon expressing $z_0$ via $T$ using \eqref{T_H} leads to 
\bea
C = \ \alpha'^{p-3}\ (g^2_{YM})^\frac{3p-13}{5-p}\ N^\frac{7-p}{5-p}\ T^\frac{9-p}{5-p}.
\eea
It is manifest that the specific heats are positive for all $p <5$. As mentioned previously, in these calculations we have not kept  track of numerical pre-factors which are unimportant for our purposes (except that they are always positive).
We can also express the specific heat in terms of the mass density of the black hole, which is $\epsilon$ as defined in eqn. (9) of \cite{Itzhaki}. This takes the form
$C \sim \epsilon^\frac{9-p}{2(7-p)}$ where we have suppressed dependence on $\alpha'$, $g_{YM}$ and $N$ as well. 

\subsubsection{Minimal Energy Quantum}

As was argued in \cite{Verlinde}, the way in which the minimal energy quantum scales with the radius (ie., towards the IR of the bulk) is an indicator of the long string phenomenon (or its absence). As we discussed in a previous section, this scaling can be computed by taking the holographic screen to be at the horizon of the black brane. The black hole mass-energy is viewed as the the excitation energy associated to all the holographic degrees of freedom on the screen. This allows us to compute the minimal energy quantum $\epsilon_{dof}$ via
\bea
\epsilon = N_{dof} \epsilon_{dof}
\eea 
The $N_{dof}$ in Einstein frame was computed above to be $N_{dof} \sim z_0^\frac{p-9}{5-p}$, and the scaling of $\epsilon$ with $z_0$ can be obtained by setting the blackening factor in the metric to zero, and using eqn. (9) of \cite{Itzhaki} together with the expressions in our Appendix \ref{DpCoord} relating coordinates. The result is 
\bea
\epsilon \sim z_0^\frac{2(p-7)}{(5-p)}.
\eea
It immediately follows that the scaling of the minimal energy quantum with the holographic radius is universal for all $p$,
\bea
\epsilon_{dof} \sim \frac{1}{z},
\eea
and that it increases towards the IR of the bulk. This is the opposite of what one expects when we have a long-string phenomenon, and it is consistent with our previous observations about IR/UV duality for Poincare Dp-branes.

\section{Comments and Open Questions}

A key suggestion of this paper is that the AdS length should be viewed as the length scale associated to the long string in a stack of $N$ D3-branes {\em after backreaction}, i.e., at strong coupling.  Repeating the representative example in \eqref{LvsN}, we can re-write:
\bea
L = (4 \pi g_s N)^{\frac{1}{4}} l_s.  \label{LvsN2} 
\eea
It is important to note that the relation \eqref{LvsN2} implicitly contains two steps. Firstly, we view it as computing the long string length-scale at strong coupling. Secondly, (and this is the part familiar from AdS/CFT) we view it as the curvature length-scale of the background. We view the equality as relating the two. We expect the relation to hold exactly in the large-$N$ limit. At weak coupling, this quantity controls closed string emission amplitudes from D-branes. At strong coupling, it again shows up as sourcing the harmonic function in the supergravity solution. Note that when we apply similar relations to sub-matrices and sub-AdS scales, and want to work in the strict large-$N$ limit, we should implicitly consider the scaling limit with $M/N$ held fixed as we send $N \rightarrow \infty$. 

At weak coupling and small number of branes, i.e., in perturbation theory, it may be\footnote{We do not take a strong stance on this point, because our results do not directly depend on the weak coupling result. An alternate perspective is that one should give a physical interpretation for the length scale in \eqref{LvsN2} even at weak coupling.} natural to expect that the length of the long string will scale linearly as $ \sim N l_s$. The expression above then suggests that this expectation gets renormalized at strong coupling. A linear long string scaling together with a conformal map prescription was used by \cite{Verlinde}, to make some statements about conical defects in AdS$_3$/CFT$_2$. Our discussion is somewhat more general and slightly different -- we have allowed long strings in stacks of branes as well, and we have taken into account backreaction. Together, this allowed us to take one more step, and propose that the AdS length {\em is} the long string length in the stack. The physics of sub-AdS length scales then emerged naturally in terms of long strings in substacks, and long string rescaling became a natural operation in the AdS geometry.

Understanding \eqref{LvsN2} as a statement about the long string length-scale, may emerge as a natural corollary of a derivation of the AdS/CFT correspondence at strong coupling. Conversely, this perspective may provide useful intuition for such a derivation. At large-$N$, there are now derivations of AdS/CFT available in the market \cite{Rajesh1, Rajesh2}. These derivations consider the system after the near-horizon limit. The CFT in a suitable free limit is demonstrated to have the spectrum and correlators of a bulk worldsheet theory in a similar limit. To make statements about the long string in the D-brane stack however, it may be useful to know the connection between the open and closed descriptions, {\em before} the near-horizon limit. In any event, it will be very interesting to understand the role of long strings in such systems. A related point is that most discussions of long strings are at weak coupling \cite{Das, SusskindFat}, and part of the challenge is to give the long string a precise definition away from perturbation theory. 

We discussed the connection between AdS curvature scale in the bulk and the backreacted long string length scale, above. It is tempting to think that curvature in the bulk should more broadly be understood in terms of the long string phenomenon. More pragmatically,  can we compute the power of $M$ in the backreacted long-string formula from stringy dynamics? Why is this a simple power law? Why is this power $1/4$ in many of the cases? Some of these questions have an understanding in terms of supergravity solutions, but it would be interesting to understand it more dynamically from string theory.

In sub-AdS scales, we mentioned that the central charge we defined {\em increases} as we go to longer length scales in the bulk. Together with IR/IR duality, this seems superficially like a violation of c-theorems and related ideas. It is difficult to resolve this question with our current level of understanding, but it seems clear that the flows one considers in sub-AdS scales should implicitly involve integrating out rows and columns of the matrix field. Similar operations have been studied in the context of matrix models in \cite{Zinn-Justin} in efforts to understand the double-scaling limit \cite{Kazakov, Douglas, Gross}. The double-scaling limit was a crucial piece in connecting matrix models and 2-dimensional string theory, and it stands to reason that something along these lines may play a role in obtaining long strings in sub-AdS scales as well. A related point is that our definition of the central charge was based on bulk metric variables -- it will be nice to have a more microscopic/stringy definition of this object. This will help clarify the precise relationship between the central charge as defined by $TT$ OPEs in field theories, and the central charge we have defined here via bulk geometry. We expect that sharpening these issues will reveal that our central charge is sufficiently different from the conventional field theory central charge, that the tension with c-theorem will evaporate. Yet another reason to think that sub-AdS scale physics of AdS$_5 \times S^5$ is controlled by a matrix model comes from the BFSS model for M-theory \cite{BFSS} -- sub-AdS physics is analogous to flat space physics (as we have alluded to many times), and it is striking that the dual of flat space in 11 dimensions is indeed a {\em matrix} model.

\section{Acknowledgments}  

This paper has its origins in collaborations with Jude Pereira. We thank Sam van Leuven for discussions and Roberto Emparan for a correspondence.  CK thanks Nirmalya Kajuri for hospitality and IIT Mandi for a stimulating environment during a key stage of this work.

\appendix

\section{Connections to Previous Work}\label{App-Verlinde}

In this Appendix we will review and contrast some of the previous ideas about sub-AdS holography that we have found useful. Our work here is a refinement of these ideas, so we will summarize them and also clarify why we believe our proposal may have some utility.

\subsection{Conformal Maps vs. Long String Re-Scaling}

We will start by reviewing aspects of \cite{Verlinde}, which together with \cite{Hanada}, was a major source of inspiration for this paper. The paper \cite{Verlinde} considered spherically symmetric timelike holographic screens in various (often maximally symmetric) spacetimes and conformal maps between them, to make statements about sub-AdS screens. We will contrast their approach to our proposal -- we view sub-AdS scales in terms of long strings in sub-stacks of D-branes and do not directly rely on conformal maps. We will see that when interpreted as a statement about sub-AdS central charges, their results match ours.  

It was proposed in \cite{Verlinde} that theories on holographic screens are identical as long as the induced metrics on them are the same. For two conformally related metrics $g$ and $\tilde g$ with $g = \Omega^2 \tilde g$, this means that a holographic theory on the screen $\mathcal{S}$ in the metric $g$ and the theory on its image screen in the metric $\Omega_\mathcal{S}^2 \Omega^{-2} g (=\Omega_\mathcal{S}^2 \tilde g)$ are identical. Here $\Omega_\mathcal{S}$ stands for the value of the conformal transformation factor {\em at} the screen\footnote{The screens in \cite{Verlinde} are radially symmetric.}. This statement was then applied to pairs of conformally related global AdS spacetimes with identical length scales $L$:
\bea
g = AdS_d \times S^{p-2} \cong  AdS_p \times S^{d-2} = \tilde g
\eea
The metric $g$ is given explicitly by eqn (A.11) of \cite{Verlinde} and $\tilde g$ by eqn (A.10). It is straightforward to check that $g = \Omega^2 \tilde g$, with 
\bea
\Omega = \frac{R}{L} = \frac{L}{r} 
\eea
where the second equality can be viewed as a coordinate transformation relating the radial coordinates in $g$ and $\tilde g$. 

The discussion of \cite{Verlinde} further invokes the notion of total number of UV degrees of freedom that lives on a holographic screen of some radius, $\mathcal{C}$. We will use $\mathcal{C}(R)$ to denote the number microscopic degrees of freedom on a screen at $R$ in the metric $g$ and $\tilde {\mathcal C}(r')$ for the same quantity on a screen at $r'$ in the metric $\Omega_{\mathcal{S}}^2 \tilde g$. It is natural that $\mathcal{C}$ is essentially the area of the screen in Planck units (of the effective lower dimensional theory!), as suggested by the Bekenstein bound. Roughly\footnote{Roughly, because comparing areas of screens in different spacetimes requires care.}, this is the horizon area of a large AdS black hole of that radius:
\bea
\mathcal{C}(R) = \frac{A(R)}{16 \pi G_d}, \ \ {\rm where} \ \ A(R) = \Omega_{d-2}R^{d-2}.
\eea 
In terms of the central charge \eqref{central}\footnote{Note that we are working with AdS$_{d}$ and not AdS$_{d+1}$ here.}, this can be written as
\bea
\mathcal{C}(R)=\frac{c}{12} \left(\frac{R}{L}\right)^{d-2}
\eea
In \cite{Verlinde} the goal then was to find an argument for $\mathcal{C}$ on constant radius sub-AdS screens, so that the result was again the corresponding area. In other words, the goal was to {\em derive} the area law for sub-AdS screens, using the area law for super-AdS screens as well as other ingredients like the conformal map. We summarize this argument in the figure below:
\begin{figure}[H]
    \centering
    \includegraphics[width=1\linewidth]{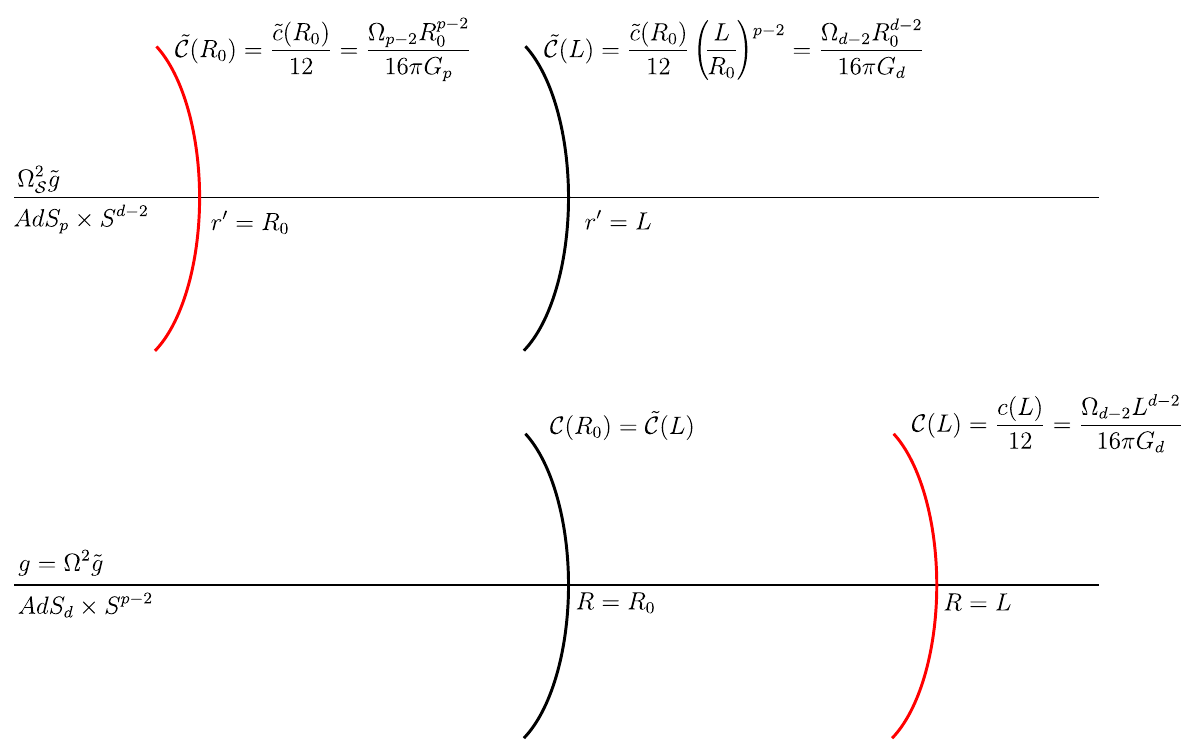}
    \caption{Summary of ``conformal-mapping-between-holographic-screens" argument of \cite{Verlinde}.}
    \label{fig:r_Vs_x_p_05}
\end{figure}  
\noindent
In the figure, the lower line denotes the radial direction of the metric $g$, while the upper line denotes that of the metric $\Omega_{\mathcal{S}}^2 \tilde g$, where $\Omega_{\mathcal{S}}=\frac{R_0}{L}$. The red screens are at the corresponding AdS radii. The black screen in the lower metric is at a sub-AdS scale and we are interested in determining the $\mathcal{C}$ associated to it. The two black screens have the same induced metric, but in the upper line, it is at a super-AdS scale. So by the proposals of \cite{Verlinde} we can compute it reliably in the upper metric and trust it as being valid for the lower black screen as well. A crucial assumption is that the higher ($d+p-2$) dimensional Newton's constant is taken to be the same for both $g$ and $\Omega_{\mathcal{S}}^2 \tilde g$, and we determine the effective lower dimensional Newton's constants $G_d$ and $G_p$ by dimensionally reducing on a sphere with radius equal to the corresponding AdS radius. For the upper metric, note that the AdS radius is {\em not} $L$, it is $R_0$:
\bea
\frac{1}{G_p}=\frac{R_0^{d-2}\Omega_{d-2}}{G_{d+p-2}}, \ \ \ \frac{1}{G_d}=\frac{L^{p-2}\Omega_{p-2}}{G_{d+p-2}}
\eea

In the upper line of the figure we have used $r'$ to denote the radial coordinate instead of $r$. This is to emphasize that the metric here is $\Omega_{\mathcal{S}}^2 \tilde g$ and  {\em not} simply  $\tilde g$. The tilde metric $\tilde g$, {\em after} this rescaling, is still an $AdS_{p} \times S^{d-2}$ metric, but the AdS radius is now $R_0$. The new radial coordinate $r'$ has been re-scaled to bring the metric to a conventional form (eqn. A.10 of \cite{Verlinde}, but with $r$ replaced by $r'$ and $L$ replaced by $R_0$).

This construction involves many moving parts. We have made an effort to streamline the presentation of \cite{Verlinde} by emphasizing the various ingredients and assumptions. In the main body of this paper, we made a minimal statement about long string re-scaling, and developed and applied it in various contexts. We will now argue that the crux of the ideas above can in fact be traced to an equivalent operation, which happens {\em not} in the conformal mapping or in the choice of the holographic screens, but in the re-scaling of $\tilde g$ to  $\Omega_{\mathcal{S}}^2 \tilde g$. This is in effect, a long-string rescaling. To see this, instead of the screen-dependent quantity $\mathcal{C}$, we can work directly with the central charge at the relevant scale:
\begin{figure}[H]
    \centering
    \includegraphics[width=1\linewidth]{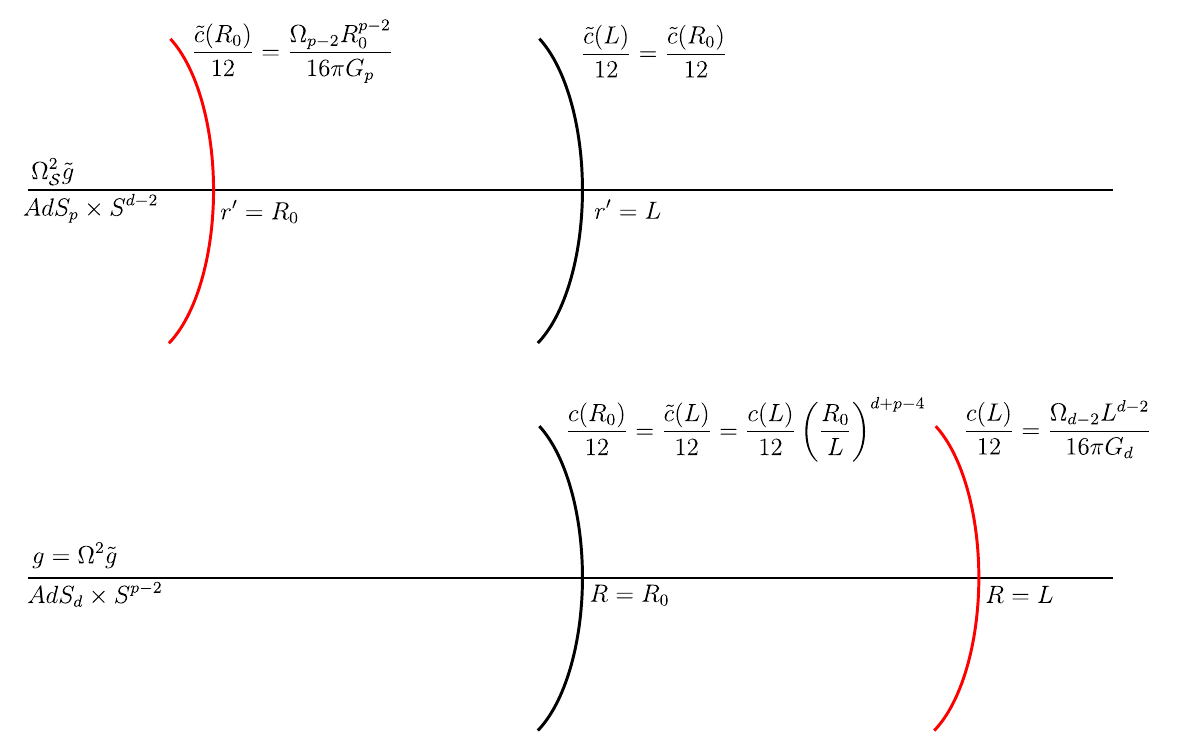}
    \caption{Summary of the argument in \cite{Verlinde}, but viewed as a statement about central charges at the relevant scales.}
    \label{central-figure}
\end{figure}
\noindent
Since the central charge stays constant at super-AdS scales, the conformal mapping argument of \cite{Verlinde} suggests that the central charge of the screen at the AdS-radius ($R_0$) in the $\Omega_{\mathcal{S}}^2 \tilde g$ metric should match with the central charge on the same metric at the scale $r'=L$, and therefore with the central charge of the sub-AdS scale $R=R_0$ in $g$. The simple calculation presented in Figure \ref{central-figure} leads to the explicit expression
\bea
\frac{c(R_0)}{12} =\Big( \frac{R_0}{L}\Big)^{d+p-4} \frac{c(L)}{12}, \label{c-rescaling}
\eea
which is the same as the one presented in \eqref{central-scaling}. It is interesting to note that the scaling in \cite{Verlinde} happens in the tilded metric, while in our long string argument there is no auxiliary metric. The reason why this works, is because the $g$ and $\tilde g$ metrics of \cite{Verlinde} have identical central charges (as is easily checked). 

\subsection{Twisted Strings, D-brane Strings and Backreaction}

In this sub-section, we will discuss a second proposal in \cite{Verlinde} which considered long strings in the symmetric product orbifold in AdS$_3$/CFT$_2$, together with the conformal mapping argument to connect it to some higher dimensional AdS geometries. In particular \cite{Verlinde} focused on ``projecting" onto a long string sector which leads to the conical defect AdS$_3/\IZ_M\times S^{d-2}$, which then gets conformally mapped to an AdS$_d \times S^1$ geometry with a radius for the $S^1$ that is hierarchically (i.e., $M$ times) smaller than the AdS$_d$ length. We will briefly discuss AdS$_3$ conical defects as an application, but it will not be our focus. We start by clarifying the distinctions between our approach and that of \cite{Verlinde}.

There are some important differences between our long string proposal and the one in \cite{Verlinde}. Firstly, the latter considers twisted sector long strings in the symmetric orbifold CFT and is specific to AdS$_3$ conical defects, and uses a conformal map to connect with certain higher dimensional AdS $\times S^1$ spaces. We are instead considering long strings in general (in particular in D-brane stacks as well). Even though often discussed in the setting of the D1-D5 CFT \cite{Das, SusskindFat, Dijkgraaf} it is worth emphasizing that long strings are a natural expectation in other holographic/black hole settings as well. In particular, the original string/black hole correspondence \cite{Susskind} relies on a long string picture. When we have a stack of D-branes, there are effective long strings that arise from bits of open strings that stretch between various branes and form closed loops. A long closed string is an analogue of a long single trace operator. Because of trace relations, the string length cannot be arbitrarily large. This perspective was emphasized in \cite{Hanada}. We view the emergence of long strings in these different settings to be a kind of long string {\em universality}\footnote{The crucial role of trace relations has also appeared recently in the context of fortuitous states \cite{Chi-MingChang, Chang2}. The idea here is that only those states that are BPS {\em after} the use of trace relations at finite-$N$ qualify as BPS black hole microstates. It is tempting to think that similar ingredients are universal for understanding even non-SUSY black hole microstates and that these states are intimately connected to long string states.}. Long string universality was crucial for us to frame our calculations within a single AdS geometry -- we did not need the conformal map nor the indirect reliance on the long strings in AdS$_3$/CFT$_2$.

(Obtaining a higher dimensional central charge in terms of the conformal map from AdS$_3$ has the property that if you start with sub-AdS$_3$ and go to super-AdS$_d$ scales, the resulting quantity is not the natural {\em higher dimensional} central charge. See eqn. (4.42) of \cite{Verlinde} and the discussion surrounding it. We wish to avoid this feature, and our prescription automatically manages to do so because we do not invoke auxiliary metrics or conformal maps to them.)

A second distinction of our proposal is that the long string rescaling in \cite{Verlinde} is a linear rescaling\footnote{In this subsection, we use $k$ to denote a general cycle size, $M$ to denote the maximal cycle in a conical defect, and $N$ to denote the power of the symmetric product orbifold.}. This means that the string in a cycle of size $k$ is (in effect) taken to be dual to a sub-AdS scale $R_0 = k l_s$ in an AdS$_3$ geometry with length scale $L= N l_s$ where $N$ is the power of the symmetric orbifold $(T^4)^N/S_N$. We have taken the target space to be $T^4$ for concreteness here, but it is not crucial for the general arguments in \cite{Verlinde}. Also, the usual AdS$_3 \times S^3$ near horizon geometry corresponds to the special case $d=5$ in the more general AdS$_3 \times S^{d-2}$ system discussed there.

The linear scaling of string length may be reasonable when the strings are weakly coupled -- but once there is backreaction in the metric description, we expect this scaling to be non-linear. In fact, in order for the long string rescaling to match correctly with the D1-D5 discussion in Section \ref{D1D5}, it is crucial that the length-scale is related non-linearly to cycle size:
\bea
R_0^4= g_6^2 \ l_s^4 \ k. \label{kR0}
\eea
This has the virtue that when $k \rightarrow N=Q_1Q_5$, we get the usual relation for the D1-D5 system \cite{Maldacena}:
\bea
L^4 = g_6^2 \ l_s^4 \ Q_1Q_5. \label{kL}
\eea
Note that if we are merely interested in describing the AdS$_3$/CFT$_2$ central charges \cite{Verlinde} in terms of the (sub- or super-AdS) length scales in the geometry, this problem is not egregious because the cycle-size dependence drops out of the final expressions. But if one wants to view the operation as a genuine long string re-scaling so that one can relate length scales to cycle sizes, backreaction cannot be ignored. In other words, the $k$ used in \cite{Verlinde} on the bulk side should really be replaced with $\sim k^{1/4}$ if one wants to view $k$ as the cycle size on the CFT side in the D1-D5 system. In other words, taking backreaction into account is what allows us to view the AdS length scale as the length of the long string.

Note that the picture we have here for $N$ in the orbifold of the D1-D5 system and the gauge group rank in other brane systems is exactly parallel. The analogue of the cycle structure in the orbifold is provided by the open strings stretched between individual branes in a stack that together give rise to a closed string  -- this is how the single trace gauge invariant is constructed from products of matrix fields \cite{Hanada}. This is the D-brane analogue of highly twisted sectors. It makes it manifest that the symmetric group is a gauge symmetry of the orbifold system in the open string description.  

We conclude with a final comment about conical defects of AdS$_3 \times S^{d-2}$ which were discussed in \cite{Verlinde}. The idea here is that for describing sub-AdS scales in an AdS$_3$ conical defect (as opposed to the AdS$_3$ vacuum), one should consider an effective long string that is smaller than the maximal one. In other words, the maximal cycle size one allows\footnote{In the discussion of \cite{Verlinde}, the notation for the maximal cycle was $M$ and the conical defects corresponded to $N < M$. We will reverse this notation because we want $N$ to be the largest stack size, across dimensions.} is no longer $N$, but $M < N$ ($M$ is taken to be a divisor of $N$). A ``long string projection" on to the $M$-cycle sector is implemented in the bulk via a quotienting of the circle in the AdS$_3$ metric by $M$. The projection is a logically distinct second step, and we will assume that it is correct in the rest of this discussion. By ensuring that the long string rescaling is done while paying heed to backreaction (like in \eqref{kR0}) instead of setting $R_0 \sim k$, this second step can be made to work in our setting as well. An immediate consequence is that the central charge of the conical defect is 
\bea
c_M(L)=\frac{c_N(L)}{M},
\eea
because the perimeter (``area") of the circle is reduced by a factor of $N$. See \cite{Bala} for a related discussion. Here we use a sub-script to denote the maximal cycle under consideration. It denotes the sector onto which the long string projection is being done\footnote{We have changed the notation from \cite{Verlinde}, who use a length-scale as the subscript.}. The sub-AdS$_3$ central charge in the conical defect is easily obtained -- it is simply the rescaling of this new central charge as in \eqref{c-rescaling}:
\bea
\frac{c_M(R_0)}{12}=\Big( \frac{R_0}{L}\Big)^{d-1} \frac{c_M(L)}{12}.
\eea
The central charges presented in \eqref{c-rescaling} (with $p=3$) will be denoted $c_N(R_0)$ and $c_N(L)$ in this notation. In \cite{Verlinde}, one further makes a conformal map of the AdS$_3/\IZ_M\times S^{d-2}$ geometry to an AdS$_d \times S^1$ geometry, which results in a compact circle that is hierarchically smaller by a factor of $M$ compared to the AdS$_d$ scale. It is an interesting question whether such scale-separated AdS compactifications with a circle are natural in string theory.  

\subsection{Pros and Cons}

As outlined in this Appendix, our proposal in this paper is a slightly modified and generalized version of the proposal in \cite{Verlinde}. Among other things, we take the AdS length scale to be the size of the long string at strong coupling. This, together with the sub-stack picture means that re-scaling of the AdS $\times M$ geometry is our fundamental operation, not the conformal map.  

Let us summarize a few places where these differences become significant:
\begin{itemize}
\item The long string mechanism in this paper is not limited to AdS$_3$ geometries and the symmetric orbifold. This is satisfying, because as already alluded to, there are natural long strings in stacks of D-branes as well. The natural gauge invariant (closed string) object that one can construct from open strings connecting various branes is a closed loop of open strings stretched between those branes. This is the D-brane realization of taking traces to construct gauge invariants. 
\item A closely related fact is that the reliance on the conformal map to connect AdS$_3$ and higher dimensional AdS$_d$ geometries, results in geometries with an extra $S^1$. This seems like a limitation on the kind of geometries that one can relate to the long string in AdS$_3$/CFT$_2$. Our approach instead works in a uniform way for all Freund-Rubin type compactifications. 
\item More generally, the conformal mapping works by switching the sphere inside the AdS with the external sphere. This means that the construction only works when the compact space is a sphere (with specific dimensionality constraints). Physically however, there is no reason why these more general geometries should not have an understanding in terms of long strings.
\item As already mentioned, the proposal in \cite{Verlinde} effectively assumes that  the long string length, scales linearly in the number of branes (or the cycle size in the symmetric orbifold language).  But even in AdS$_3$ settings (like in the AdS$_3 \times S^3$ geometry arising in the D1-D5 system) such a prescription would lead to tension with how the AdS length is related to the cycle size in the fully backreacted geometry. This is simply the observation that the $L$ and $N$ in \eqref{LvsN} are not related linearly. We instead suggested that the long string length can be directly compared to the AdS length scale if one includes backreaction. Note that some non-linearity is natural to expect as the branes start becoming heavy.
\item A closely related fact is that our alternate scaling naturally suggests that  the thermodynamics of small black holes comes out right, when one works with a sub-stack of $M$ branes instead of the full $N$ stack. It provides an explanation for the ``smallest large black hole" picture suggested in \cite{Hanada} for a small black hole. 
\item The conformal mapping proposal of \cite{Verlinde} maps super and sub-AdS screens in AdS$_d \times S^{1}$ and AdS$_3 \times S^{d-2}$ geometries.  But when one does the map twice to go back to super-AdS scales, the resulting central charge in AdS$_d$ does not have a simple AdS$_d$ interpretation. In particular, it is not constant. By dispensing with the conformal map entirely, we are able to avoid this feature. %This is not to suggest that the conformal map prescription is incorrect. We simply wish to keep our ingredients to a minimum and {\em not} have to deal with conceptual questions that are hard to resolve at this stage.
\item The arguments of \cite{Verlinde} were made using constant radius screens in global AdS. These screens are natural proxies for black hole horizons at super-AdS scales, but they are somewhat unnatural at sub-AdS scales. The reason is that black holes with such horizons at sub-AdS scales suffer from a Gregory-Laflamme instability. Our approach avoids such sub-AdS screens. 
\item On the flip side, since our approach does not use a conformal map, we have nothing to say about Minkowski or de Sitter compactifications which were alluded to in \cite{Verlinde}.
\end{itemize}

%The conformal map seems to need two spacetimes. Are they living in the same theory? The higher dimensional G reduces to different lower dimensional G's depending on the screen. This looks weird, but is it really? (But does our proposal also effectively lead to that?)

\section{Gregory-Laflamme Instability and the Near-Horizon Limit}

Roughly, GL instability arises when a horizon has long extended directions, together with a thin cross-section. A planar AdS-BH has an infinite cross section (it is a wall that extends in all the directions transverse to the radial direction). Therefore, it is intuitive that it does not satisfy the GL mode existence condition, and is therefore stable\footnote{This is a little too quick, because the AdS length scale also has a role to play. In flat space, such wall horizons do not exist (for example). But these subtleties are not important for our discussion.}. Note that this is not true in flat space because black branes are not real ``walls" of the form $\IR^{m+n}$. Rather, they are of the form $\IR^m \times S^n$, and therefore flat space black branes can have small cross sections along $S^n$ while being extended along $\IR^m$. So they are GL unstable.

One way to understand the near horizon limit is that one is zooming in near a flat space black brane, so that the $\IR^4 \times S^5$ transverse to the radial direction becomes an $\IR^4$ (the $S^5$ becomes an overall constant factor multiplying the entire space). Because of this, if one is sufficiently close to the stack, the emergent geometry loses the GL instability mechanism. In this case, the temperature of the brane has an interpretation as the CFT temperature. For the small black hole localized on the $S^5$ on the other hand, the implicit assumption is that there is a small hierarchy between the stack length scales, $R \lesssim L$. This is what allows one to resolve the D3 sources localized on the $S^5$ in the asymptotically AdS$_5 \times S^5$ geometry, after the decoupling limit. If these branes are heated up, they are still aware of the local $\IR^4 \times S^5$ structure, and are essentially like flat space branes as far as the GL instability is considered. Therefore we can expect them to clump and form black holes (rather than stay as black branes). In other words, empty AdS$_5 \times S^5$ cannot have {\em localized} stable black branes living in it, for the same reason that flat space cannot. 

A different-sounding, but fundamentally identical idea is to say this is as follows -- the reason why planar-AdS black branes do not have an instability is because we think of them as being heated up after the near horizon limit, and we are raising the temperature of the whole CFT. Relatedly, the AdS length gives a natural length-scale, which is shorter than the long wavelength GL mode. Instead if we are heating up a D3 brane stack in some {\em localized} region (before or after the near horizon limit  -- ie., flat space or AdS), we can expect a GL instability.

Let us also emphasize that all these discussions about instabilities are at suitably finite temperatures, when supersymmetry is broken.

\section{Coordinates for Decoupled Dp-branes}
\label{DpCoord}

In this appendix, we present a few coordinates for the near horizon Dp-brane geometries that we found useful. 

The general Dp-brane metric (after the decoupling limit) can be written in a manifestly conformally AdS form \cite{Magoo}: 
\bea \label{MagooDp}
ds^2=\alpha' \Big(\frac{2}{5-p}\Big)^\frac{7-p}{5-p}(c_p g^2_{YM}N)^\frac{1}{5-p}z^\frac{3-p}{5-p}\left[\frac{-dt^2+dx_p^2+dz^2}{z^2}+\Big(\frac{5-p}{2}\Big)^2d\Omega_{8-p}^2\right].
\eea
This metric is often convenient for describing $p \le 3$ branes. For $ p \ge 3$ we find the form 
\bea
ds^2=\alpha' \Big(\frac{2}{5-p}\Big)^2(c_p g^2_{YM}N)^\frac{1}{2}{\tilde z}^\frac{3-p}{5-p}\left[\frac{-dt^2+dx_p^2+d{\tilde z}^2}{{\tilde z}^2}+\Big(\frac{5-p}{2}\Big)^2d\Omega_{8-p}^2\right]
\eea
useful, which involves a slight rescaling of the $z$ coordinate
\bea
\tilde z \equiv  \frac{5-p}{2}\frac{z}{\sqrt{{c_p}g_{YM}^2 N}}= \frac{1}{U^{\frac{5-p}{2}}}.
\eea
Here the last equality is the definition of the $z$-coordinate as given in \cite{Magoo}. $U = r/\alpha'$ is the radial coordinate held fixed in the near-horizon limit and $c_p$ is a $p$-dependent numerical constant which will not play any important role in our discussion. 

Both the above forms manifest the fact that the geometry is conformal to AdS$_{p+2} \times S^{8-p}$. In terms of the $U$ coordinate, the regime where the supergravity description is valid can be written as \cite{Itzhaki}:
\bea	
1 \ll g^2_{YM}NU^{p-3} \ll N^{\frac{4}{7-p}}.
\eea
The quantity $g^2_{eff} \sim g^2_{YM}NU^{p-3}$ is the  dimensionless (but scale-dependent in $p \neq 3$) effective 't Hooft coupling, which is what we will often mean by the gauge coupling. 
For $p \le 3$ the range above where the 10 d supergravity approximation is valid becomes 
\bea
\frac{1}{(g^2_{YM}N)^{\frac{1}{3-p}}} \ll z \ll \frac{N^{\frac{1}{7-p}}}{(g^2_{YM})^{\frac{1}{3-p}}} \label{sugra-p<3}
\eea
and for $p \ge 3$ we have
\bea
\frac{(g^2_{YM})^\frac{5-p}{2(p-3)}}{N^\frac{5-p}{2(7-p)}} \ll \tilde z \ll (g^2_{YM} N)^\frac{5-p}{2(p-3)}. \label{sugra-p>3}
\eea
We have chosen these coordinates so that this range spans the positive real half-line in the large-$N$ limit for any $p$. Note that without a judicious choice of such a coordinate (e.g., if we were to work with $U$ or $z$ for $p > 3$), both ends of the inequalities either go to zero or infinity which can sometimes be inconvenient for thinking about them.

The coordinates above have dimensions. But a simple dimensionless coordinate that we can introduce is
\bea
\hat z \equiv (g^2_{YM} N)^{\frac{1}{3-p}}\  z .
\eea
After a suitable re-scaling of the $t$ and $x_p$ coordinates, this leads to the metric
\bea \label{BlBrane}
ds^2=\alpha' \Big(\frac{2}{5-p}\Big)^\frac{7-p}{5-p}c_p^\frac{1}{5-p}\hat z^\frac{3-p}{5-p}\left[ds^2_{BH}+\Big(\frac{5-p}{2}\Big)^2d\Omega_{8-p}^2\right],
\eea
where
\bea \label{BlackPart}
ds^2_{BH}\equiv \frac{1}{\hat z^2}\left[-\left(1-(\hat z/ \hat z_0)^{\frac{2(7-p)}{5-p}}\right)dt^2+dx_p^2+\frac{d\hat z^2}{\left(1-(\hat z/ \hat z_0)^{\frac{2(7-p)}{5-p}}\right)}\right].
\eea
Here we have put the branes at finite temperature by introducing the horizon at $\hat z = \hat z_0$. Setting  $z_0=\infty$ leads to the empty near-horizon geometries, again in the manifestly conformal to AdS form. In the zero temperature limit, we have
\bea 
ds^2=\alpha' \Big(\frac{2}{5-p}\Big)^\frac{7-p}{5-p}(c_p )^\frac{1}{5-p}\hat z^\frac{3-p}{5-p}\left[\frac{-dt^2+dx_p^2+d\hat z^2}{\hat z^2}+\Big(\frac{5-p}{2}\Big)^2d\Omega_{8-p}^2\right].
\eea
The 10 d supergravity description is well-defined in the regime
\bea
1 \ll \hat z^{\frac{2(3-p)}{5-p}} \ll N^{\frac{4}{7-p}}.
\eea
Note that in all the cases we consider (i.e., $p<5$) the IR of the bulk corresponds to all the coordinates $(z, \tilde z, \hat z) \rightarrow 0$ while $U \rightarrow \infty$.

\end{document}